\documentclass[preprint,twocolumn]{aastex63}
\usepackage{float, bm, graphicx, amsmath, morefloats}
\usepackage[caption=false]{subfig}
\shorttitle{SN 2023zaw}
\shortauthors{Das et al.}

\usepackage{float,graphicx,amsmath,tabularx,booktabs,natbib,threeparttable}
\usepackage{float, bm, graphicx, amsmath, morefloats}
\usepackage[caption=false]{subfig}
\usepackage{float,graphicx,amsmath,tabularx,booktabs,natbib,threeparttable}
\usepackage{subfig}
\usepackage[para]{footmisc}
\usepackage{tablefootnote}

\usepackage{graphicx}
\usepackage[caption=false]{subfig}
\usepackage{lipsum} % just for the example
\usepackage{placeins}

\newcommand{\lpipe}{\textsc{lpipe}}
\newcommand{\emcee}{\textsc{emcee}}
\newcommand{\mesa}{\textsc{mesa}}
\newcommand{\snec}{\textsc{snec}}

\usepackage{comment}
\definecolor{dark-red}{rgb}{0.4,0.15,0.15}
\definecolor{dark-blue}{rgb}{0.15,0.15,0.4}
\definecolor{medium-blue}{rgb}{0,0,0.5}
\hypersetup{
    colorlinks, linkcolor={dark-red},
    citecolor={dark-blue}, urlcolor={medium-blue}
}

\newcommand{\beqa}{\begin{eqnarray}} 
\newcommand{\eeqa}{\end{eqnarray}}

\newcommand{\bsub}{\begin{subequations}}
\newcommand{\esub}{\end{subequations}}
\newcommand{\beal}{\begin{align}}
\newcommand{\ealn}{\end{align}}

\newcommand{\Nif}{$\rm ^{56}Ni$}

\newcommand{\msun}{M$_{\sun}$}

\newcommand{\Msun}{{\ensuremath{\mathrm{M}_{\odot}}}}
\newcommand{\Rsun}{{\ensuremath{\mathrm{R}_{\odot}}}}

\graphicspath{{./}{figures/}}

\begin{document}

\title{\vspace{0.5cm} 
SN 2023zaw: an ultra-stripped, nickel-poor supernova from a low-mass progenitor

%\textcolor{black}{(suggest titles!)} \\or\\
%SN 2023zaw: a stripped-envelope supernova with the lowest nickel mass\\or\\
}

\author[0000-0001-8372-997X]{Kaustav~K.~Das}\thanks{E-mail: kdas@astro.caltech.edu}
\affiliation{Cahill Center for Astrophysics, California Institute of Technology, MC 249-17, 
1200 E California Boulevard, Pasadena, CA, 91125, USA}

\author[0000-0002-4223-103X]{Christoffer Fremling}
\affil{Caltech Optical Observatories, California Institute of Technology, Pasadena, CA 91125, USA}

\author[0000-0002-5619-4938]{Mansi~M.~Kasliwal}
\affiliation{Cahill Center for Astrophysics, 
California Institute of Technology, MC 249-17, 
1200 E California Boulevard, Pasadena, CA, 91125, USA}

\author[0000-0001-6797-1889]{Steve Schulze}
\affiliation{Center for Interdisciplinary Exploration and Research in Astrophysics (CIERA), Northwestern University, 1800 Sherman Ave, Evanston, IL 60201, USA}

\author[0000-0003-1546-6615]{Jesper Sollerman}
\affiliation{The Oskar Klein Centre, Department of Astronomy, Stockholm University, AlbaNova, SE-10691 Stockholm, Sweden}

\author[0000-0003-2758-159X]{Viraj Karambelkar}
\affiliation{Cahill Center for Astrophysics, 
California Institute of Technology, MC 249-17, 
1200 E California Boulevard, Pasadena, CA, 91125, USA}

\author[0000-0003-4725-4481]{Sam Rose}
\affiliation{Cahill Center for Astrophysics, 
California Institute of Technology, MC 249-17, 
1200 E California Boulevard, Pasadena, CA, 91125, USA}

%reducers & PIs

\author[0000-0003-3768-7515]{Shreya Anand}
\affiliation{Cahill Center for Astrophysics, 
California Institute of Technology, MC 249-17, 
1200 E California Boulevard, Pasadena, CA, 91125, USA}

\author[0000-0002-8977-1498]{Igor Andreoni}
\affiliation{Joint Space-Science Institute, University of Maryland, College Park, MD 20742, USA}
\affiliation{Department of Astronomy, University of Maryland, College Park, MD 20742, USA}
\affiliation{Astrophysics Science Division, NASA Goddard Space Flight Center, Mail Code 661, Greenbelt, MD 20771, USA}

\author{Marie Aubert} \affiliation{Université Clermont-Auvergne, CNRS, LPCA, 63000 Clermont-Ferrand, France}

\author[0000-0003-1325-6235]{Sean J. Brennan} \affiliation{The Oskar Klein Centre, Department of Astronomy, Stockholm University, AlbaNova, SE-10691 Stockholm, Sweden}

\author{S. Bradley Cenko}
\affiliation{Joint Space-Science Institute, University of Maryland, College Park, MD 20742, USA}
\affiliation{Astrophysics Science Division, NASA Goddard Space Flight Center, MC 661, Greenbelt, MD 20771, USA}

\author[0000-0002-8262-2924]{Michael W. Coughlin}
\affiliation{School of Physics and Astronomy, University of Minnesota, Minneapolis, MN 55455, USA}

\author[0000-0002-9700-0036]{B. O'Connor}
\affiliation{McWilliams Center for Cosmology and Astrophysics, Department of Physics, Carnegie Mellon University, Pittsburgh, PA 15213, USA}

\author[0000-0002-8989-0542]{Kishalay De}
\affiliation{MIT-Kavli Institute for Astrophysics and Space Research
77 Massachusetts Ave. Cambridge, MA 02139, USA}

\author[0000-0002-4544-0750]{Jim Fuller}
\affiliation{Cahill Center for Astrophysics, California Institute of Technology, MC 249-17, 
1200 E California Boulevard, Pasadena, CA, 91125, USA}

\author[0000-0002-3168-0139]{Matthew Graham}
\affiliation{Cahill Center for Astrophysics, 
California Institute of Technology, MC 249-17, 
1200 E California Boulevard, Pasadena, CA, 91125, USA}

\author{Erica Hammerstein}
\affiliation{Department of Astronomy, University of Maryland, College Park, MD 20742, USA}

\author[0000-0003-4287-4577]{Annastasia Haynie}
\affiliation{Department of Physics and Astronomy, University of Southern California, Los Angeles, CA 90089, USA,}
\affiliation{The Observatories of the Carnegie Institution for Science, 813 Santa Barbara St., Pasadena, CA 91101, USA}

\author{K-Ryan Hinds}
\affiliation{Astrophysics Research Institute, Liverpool John Moores University, IC2,  Liverpool L3 5RF, UK}

\author{Io Kleiser}
\affiliation{Cahill Center for Astrophysics, 
California Institute of Technology, MC 249-17, 
1200 E California Boulevard, Pasadena, CA, 91125, USA}

\author[0000-0001-5390-8563]{S.~R.~Kulkarni}
\affiliation{Cahill Center for Astrophysics, 
California Institute of Technology, MC 249-17, 
1200 E California Boulevard, Pasadena, CA, 91125, USA}

\author{Zeren Lin}
\affiliation{Cahill Center for Astrophysics, 
California Institute of Technology, MC 249-17, 
1200 E California Boulevard, Pasadena, CA, 91125, USA}

\author[0000-0002-7866-4531]{Chang~Liu}
\affil{Department of Physics and Astronomy, Northwestern University, 2145 Sheridan Rd, Evanston, IL 60208, USA}
\affil{Center for Interdisciplinary Exploration and Research in Astrophysics (CIERA), Northwestern University, 1800 Sherman Ave, Evanston, IL 60201, USA}

\author[0000-0003-2242-0244]{Ashish~A.~Mahabal}
\affiliation{Division of Physics, Mathematics and Astronomy, California Institute of Technology, Pasadena, CA 91125, USA}
\affiliation{Center for Data Driven Discovery, California Institute of Technology, Pasadena, CA 91125, USA}

\author[0000-0002-8650-1644]{Christopher Martin}
\affiliation{Cahill Center for Astrophysics, 
California Institute of Technology, MC 249-17, 
1200 E California Boulevard, Pasadena, CA, 91125, USA}

\author[0000-0001-9515-478X]{Adam A.~Miller}
\affil{Department of Physics and Astronomy, Northwestern University, 2145 Sheridan Rd, Evanston, IL 60208, USA}
\affil{Center for Interdisciplinary Exploration and Research in Astrophysics (CIERA), Northwestern University, 1800 Sherman Ave, Evanston, IL 60201, USA}

\author{James D. Neill}
\affiliation{Cahill Center for Astrophysics, 
California Institute of Technology, MC 249-17, 
1200 E California Boulevard, Pasadena, CA, 91125, USA}

\author[0000-0001-8472-1996]{Daniel A.~Perley}
\affiliation{Astrophysics Research Institute, Liverpool John Moores University, IC2,  Liverpool L3 5RF, UK}

\author[0000-0002-8041-8559]{Priscila J.~Pessi} \affiliation{The Oskar Klein Centre, Department of Astronomy, Stockholm University, AlbaNova, SE-10691 Stockholm, Sweden}

\author[0000-0001-5847-7934]{Nikolaus Z.~Prusinski}
\affiliation{Cahill Center for Astrophysics, 
California Institute of Technology, MC 249-17, 
1200 E California Boulevard, Pasadena, CA, 91125, USA}

\author{Josiah Purdum}
\affiliation{Caltech Optical Observatories, California Institute of Technology, Pasadena, CA 91125, USA}

\author[0000-0002-7252-5485]{Vikram Ravi}
\affiliation{Cahill Center for Astrophysics, 
California Institute of Technology, MC 249-17, 
1200 E California Boulevard, Pasadena, CA, 91125, USA}

\author[0000-0001-7648-4142]{Ben Rusholme}
\affiliation{Cahill Center for Astrophysics, 
California Institute of Technology, MC 249-17, 
1200 E California Boulevard, Pasadena, CA, 91125, USA}

\author[0000-0003-2872-5153]{Samantha Wu}
\affiliation{Cahill Center for Astrophysics, California Institute of Technology, MC 249-17, 
1200 E California Boulevard, Pasadena, CA, 91125, USA}

\author[0000-0002-9998-6732]{Avery Wold}
\affiliation{Cahill Center for Astrophysics, 
California Institute of Technology, MC 249-17, 
1200 E California Boulevard, Pasadena, CA, 91125, USA}

\author[0000-0003-1710-9339]{Lin~Yan}
\affil{Caltech Optical Observatories, California Institute of Technology, Pasadena, CA 91125, USA}

%\author{Friends}
%\affiliation{xxxx}

%\email{Email: kdas@astro.caltech.edu}

%builders

%\end{comment}

%\author{Builders + Friends}

\begin{abstract} 
%We present the discovery, observations, and analysis of SN 2023zaw $-$ a sub-luminous and fast-evolving supernova (SN) at a distance of 45 Mpc. It is classified as a Type Ib SN based on the photospheric spectra. The late-time spectra show prominent narrow \ion{He}{1} emission lines, indicative of interaction with He-rich circumstellar material. The lightcurve peaks at a $r$-band absolute magnitude of $-$16.7 mag. It is the fastest evolving stripped-envelope SN discovered so far $-$ with a rise-time of 1.8 days, a fade time of 3.1 days, and an initial decline rate of $\sim$0.3 mag $\mathrm{day^{-1}}$. SN 2023zaw is located in a star-forming galaxy with a local star formation rate of \textcolor{black}{xxx}. Based on analytical and radiation-hydrodynamics modeling with \mesa\ and \snec, we obtain an envelope mass of $\sim$0.2 \msun\ and a radius of $\sim$50 \Rsun. The estimated nickel mass is $\sim$0.003 \Msun, the lowest in any stripped-envelope SN discovered so far. The estimated ejecta mass ($\sim$0.5 \Msun), and other explosion properties make it the nearest ultra-stripped SN candidate.  The explosion properties are consistent with a low-mass progenitor in a binary system with an initial mass of less than 12 \Msun. SN 2023zaw underscores the existence of an undiscovered population of extremely low nickel mass ($< 0.005 $\msun) stripped-envelope supernovae, which can be explored with deep and high-cadence transient surveys.

We present SN 2023zaw $-$ a sub-luminous ($\mathrm{M_r} = -16.7$ mag) and rapidly-evolving supernova ($\mathrm{t_{1/2,r}} = 4.9$ days), with the lowest nickel mass ($\approx0.002$ \Msun) measured among all stripped-envelope supernovae discovered to date. The photospheric spectra are dominated by broad \ion{He}{1} and Ca NIR emission lines with velocities of $\sim10\ 000 - 12\ 000$ $\mathrm{km\ s^{-1}}$. The late-time spectra show prominent narrow \ion{He}{1} emission lines at $\sim$1000$\ \mathrm{km\ s^{-1}}$, indicative of interaction with He-rich circumstellar material. SN 2023zaw is located in the spiral arm of a star-forming galaxy. We perform radiation-hydrodynamical and analytical modeling of the lightcurve by fitting with a combination of shock-cooling emission and nickel decay. The progenitor has a best-fit envelope mass of $\approx0.2$ \msun\ and an envelope radius of $\approx50$ \Rsun. The extremely low nickel mass and low ejecta mass ($\approx0.5$ \Msun) suggest an ultra-stripped SN, which originates from a mass-losing low mass He-star (ZAMS mass $<$ 10 \Msun) in a close binary system. This is a channel to form double neutron star systems, whose merger is detectable with LIGO. SN 2023zaw underscores the existence of a previously undiscovered population of extremely low nickel mass ($< 0.005$ \msun) stripped-envelope supernovae, which can be explored with deep and high-cadence transient surveys.

\end{abstract}

\keywords{supernovae: general --~ supernovae: individual (SN 2023zaw, ZTF23absbqun) ~--~
          stars: massive ~--~
          stars: mass-loss ~--~
          stars: neutron\\
         }

\section{Introduction} \label{sec:intro}

Modern wide-field and high-cadence transient surveys such as the Zwicky Transient Facility \citep[ZTF;][]{Bellm2019, Graham2019} and the Asteroid Terrestrial-impact Last Alert System \citep[ATLAS; ][]{Tonry2018, Smith2020} have expanded the discovery space of unusual supernovae (SNe). The class of rapidly-evolving, faint SNe constitutes such a population of peculiar SNe that evolve on the shortest timescales of a few days and have a peak absolute magnitude fainter than $-17$ mag. 
There are only a handful of well-studied examples in the literature $-$ SN 2005ek \citep{Drout2013}, SN 2010X \citep{Kasliwal2010}, iPTF14gqr \citep{De2018a}, SN 2018kzr \citep{McBrien2019}, SN 2019dge \citep{Yao2020}, SN 2019wxt \citep{Shivkumar2023, Agudo2023}, SN 2019bkc \citep{Prentice2020, Chen2020}, and SN 2022agco \citep{Yan2023}. 

Despite advances in understanding the photometric and spectroscopic diversity, their progenitors and powering mechanisms remain unknown. The various theoretical scenarios include ultra-stripped core-collapse SNe of massive stars that lead to binary neutron star systems \citep{Tauris2013, Tauris2015}, non-terminal thermonuclear detonations of a helium shell on the surface of a white dwarf called a `.Ia' SN \citep{Bildsten2007, Shen2010}, shock-cooling of H-poor stars with an extended envelope \citep{Kleiser2014, Kleiser2018}, fallback in a core-collapse SN \citep{Zhang2008, Moriya2010}, accretion-induced collapse of a white dwarf \citep{Dessart2007}, electron-capture SN \citep{Moriya2016}, and neutron-star white-dwarf mergers \citep{Margalit2016}.

In this paper, we present SN 2023zaw -- a Type Ib SN with the lowest nickel mass measured among all stripped-envelope SNe discovered to date. The late-time spectra are dominated by narrow \ion{He}{1} emission lines, suggestive of interaction with He-rich circumstellar material (CSM). SN 2023zaw exhibits the most rapid rise and fade time among all stripped-envelope SNe in the literature. Additionally, it is the nearest SN among all fast and faint SNe in the literature. The photometric and spectroscopic properties suggest that it is an ultra-stripped SN, originating from a progenitor with an initial mass $<$ 10 \Msun, which likely formed a neutron star.  

%Overall, the low luminosity, fast-evolving lightcurve and unique spectra of SN\,2023zaw is unlike any supernova discovered so far.

\section{Discovery and follow-up observations}
\label{sec:data}
\subsection{Discovery}

SN 2023zaw (ZTF23absbqun) was discovered with the ZTF camera \citep{Dekany20}, which is mounted on the 48-inch Samuel Oschin Telescope at the Palomar Observatory. It was first detected at $\alpha= 04^\textrm{h}29^\textrm{m}20.235^\textrm{s}$,  $\delta=+70^{\circ}25'37.52''$ (J2000) on December 7, 2023 at 05:34:06 UTC and reported to the Transient Name Server\footnote{https://www.wis-tns.org/} \citep{Sollerman2023}. At the time of discovery, the AB apparent magnitude in the $g-$band was 19.34 $\pm$ 0.15 mag. The transient stood out due to its faintness and extremely fast evolution (see Figure \ref{fig:timescale}). It was also flagged as a fast transient candidate by the ZTFReST framework for kilonova and fast transient discovery \citep{Andreoni2021} and was saved as part of the magnitude-limited \citep[Bright Transient Survey;][]{Fremling2020} and volume-limited \citep[Census of the Local Universe;][]{De2020} surveys of ZTF. %TNS for the transient \citep{Karambelkar2023aa}. 

%\subsection{Host galaxy}

\begin{figure*}
    \centering
    \includegraphics[width=5.9cm]{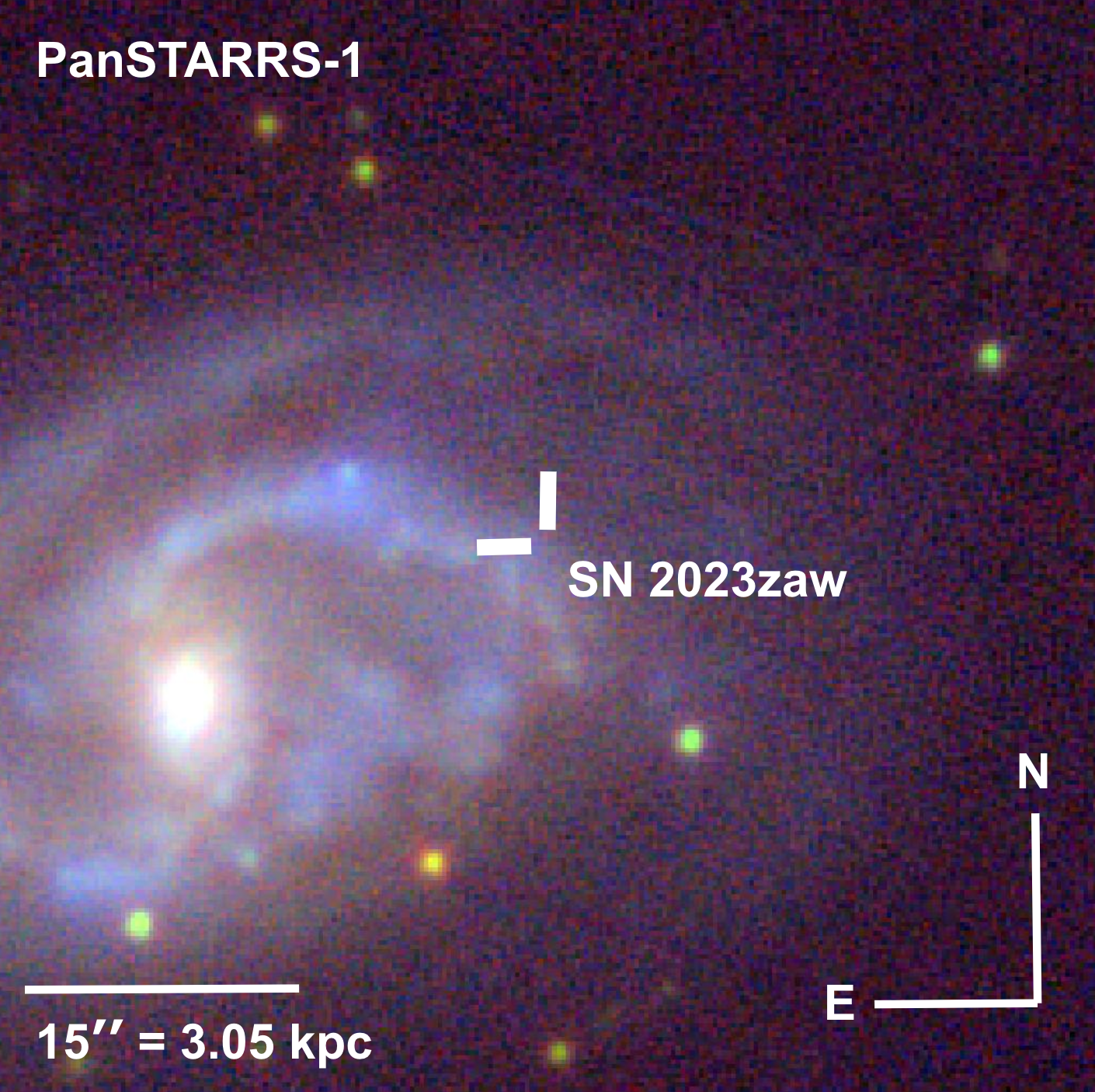}\includegraphics[width=10.2cm]{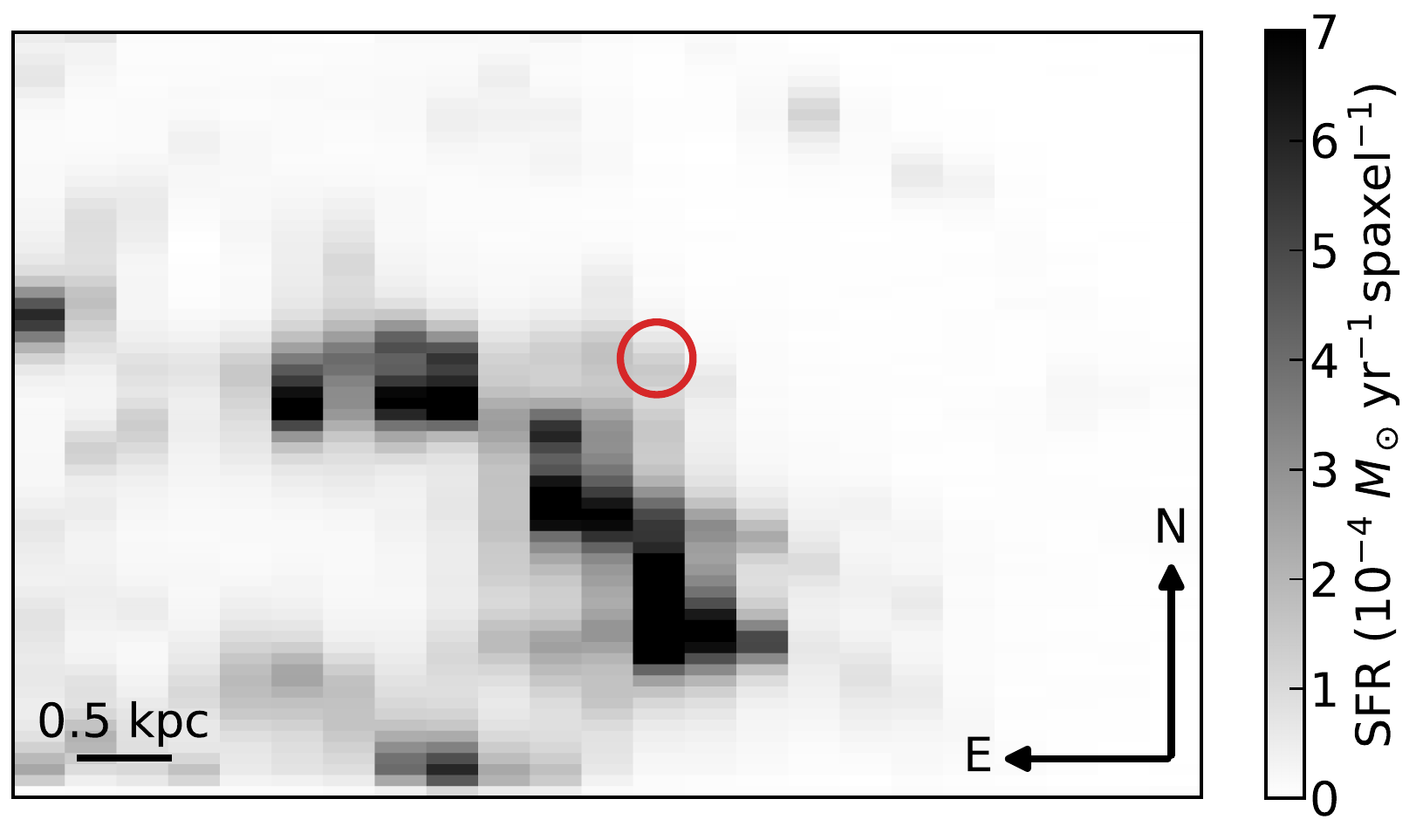}
    \caption{\textit{Left:} Pan-STARRS image of the host-galaxy of SN 2023zaw, UGC 03048 in the $grizy$ filters. The location of the transient is indicated by the white cross at the center. \textit{Right:} Star-formation rate (SFR) map of the environment around SN\,2023zaw (see Section \ref{sec:host}). SN\,2023zaw (position marked by the red circle) exploded in a star-forming region of the host galaxy close to regions of more vigorously star-forming regions. The image has a size of $4.1\times6.7$~kpc. The SFR scale is not corrected for attenuation.}
    \label{fig:galaxy}
\end{figure*}

SN 2023zaw is located in the UGC 03048 galaxy at a redshift of $z = 0.0101$ \citep{Springob2005}. It is located on the edge of one of its spiral arms (see Figure \ref{fig:galaxy}). The angular separation from the nucleus of the galaxy is 21.44$''$, which corresponds to a physical separation of 4.36 kpc. We correct for the Virgo, Great Attractor, and Shapley supercluster infall  \citep{Mould2000} based on the NASA Extragalactic Database object page (NED)\footnote{https://ned.ipac.caltech.edu/} for UGC 03048. We adopt a Hubble-flow distance of 43.9 $\pm$ 3.1 Mpc, which corresponds to a distance modulus of $33.21 \pm 0.15$ mag. 

\begin{figure*}
    \centering
    \includegraphics[width=10.5cm]{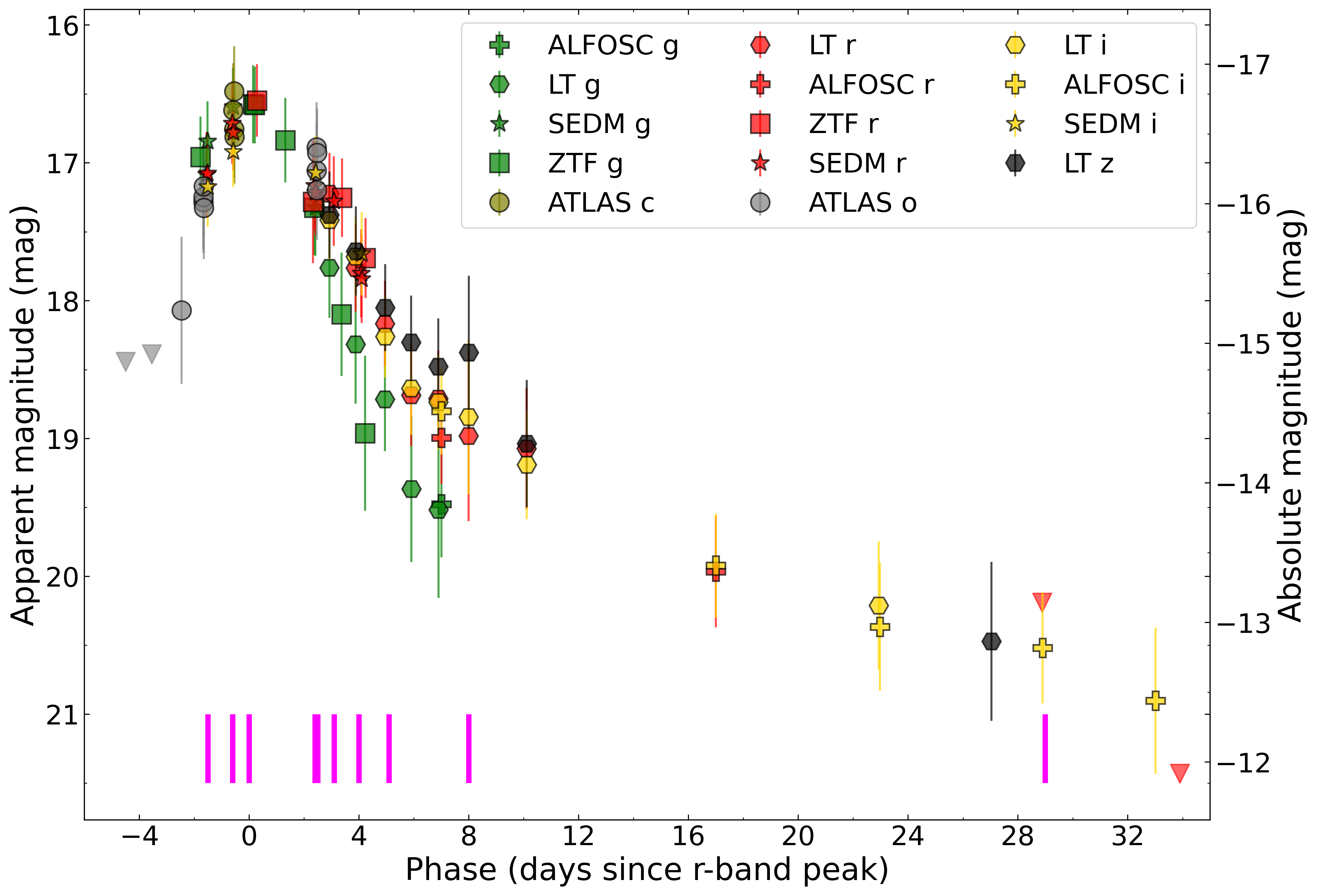}\includegraphics[width=8.5cm]{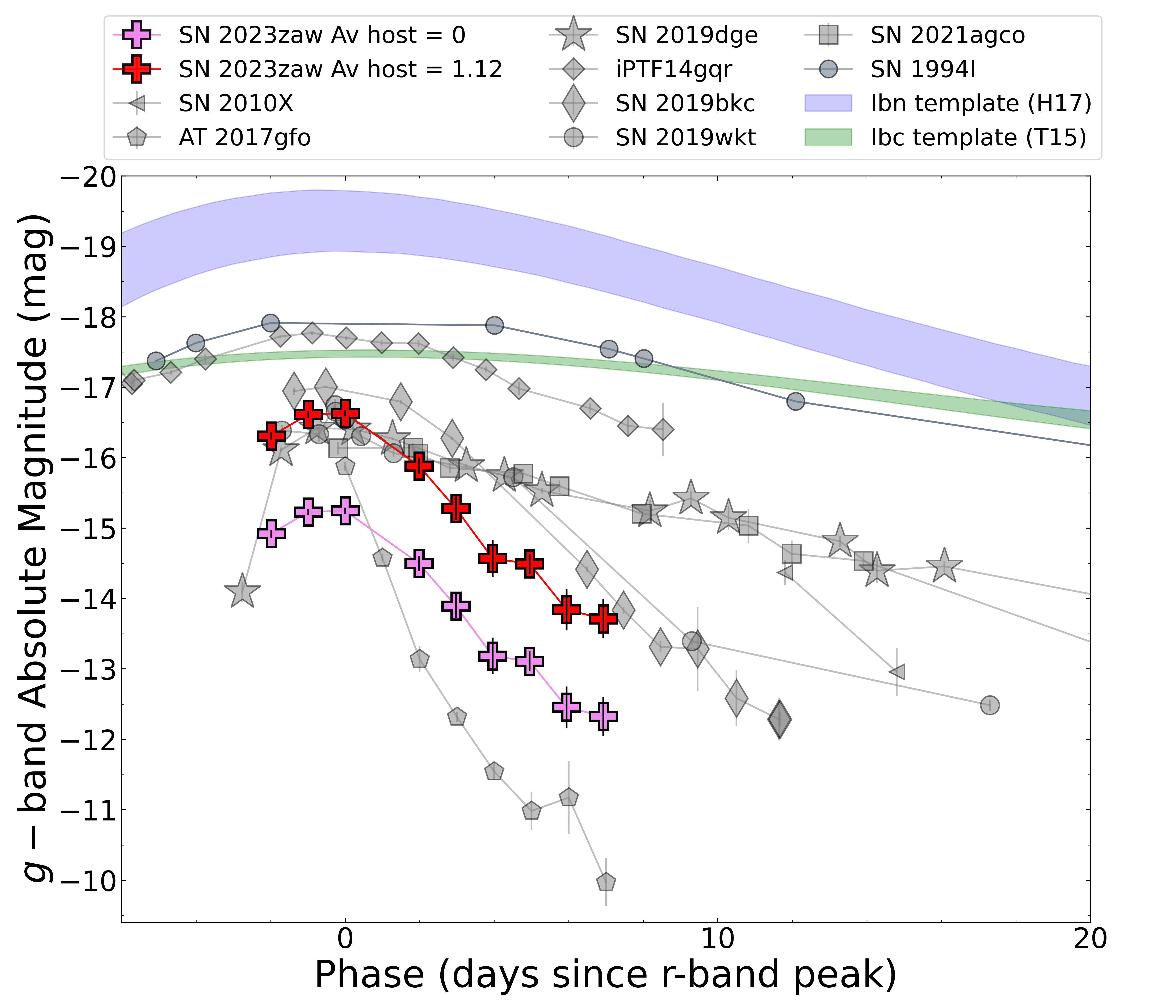}
    \caption{\textit{Left:} The multiband lightcurve collage of SN 2023zaw. The $x$-axis shows rest-frame days since the $r$-band peak (MJD = 60287.7). The photometry data has been corrected for Milky Way ($\mathrm{A_{V,MW}} =$ 0.77 mag) and host extinction ($\mathrm{A_{V,host}} =$ 1.12 mag) as described in Section \ref{sec:extinction}. \textcolor{black}{The vertical magenta lines represent epochs of spectral observations}. \textit{Right:} The $g$-band lightcurve of SN 2023zaw compared with other fast-evolving transients in the literature $-$ fast-declining Type I SNe $-$ SN 2010X \citep{Kasliwal2010}, SN 2019bkc \citep{Prentice2020, Chen2020}, the kilonova AT 2017gfo \citep{Abbott2017}, ultra-stripped SN candidates $-$ SN 2019dge \citep{Yao2020}, iPTF14gqr \citep{De2018c}, SN 2019wxt \citep{Shivkumar2023, Agudo2023}, SN 2022agco \citep{Yan2023}. We also compare with Type Ibc and Type Ibn SN lightcurve templates from \citet{Taddia2015} and \citet{Hosseinzadeh2017}, respectively.}
    \label{fig:phot_all}
\end{figure*}

\subsection{Optical photometry}
\label{section:photdata}

We obtained multiple epochs of $g$, $r$ and $i-$band photometry with the ZTF camera. The images were reduced using the ZTF image analysis pipeline \citep{Masci2019}. We also perform forced photometry at the location of the transient for the ZTF images in the $gri$ bands. We obtained $c$ and $o$-band photometry from the forced-photometry service of the Asteroid Terrestrial-impact Last Alert System \citep[ATLAS;][]{Tonry2018, Smith2020, Shingles2021}. We took regular cadence multiband photometry in the $g$, $r$, $i$ and $z$ bands with the Optical Imager (IO:O) at the 2.0 m robotic Liverpool Telescope \citep{Steele2004}. In addition, photometric data was obtained with the Rainbow Camera on the automated Palomar 60-inch telescope \citep[P60;][]{Cenko2006}. LT and P60 images were processed using the FPipe \citep{Fremling2016} image subtraction pipeline with PanSTARRS \citep[PS1;][]{Chambers2016} reference images. The Alhambra
Faint Object Spectrograph and Camera (ALFOSC) at
the 2.5m Nordic Optical Telescope was used to obtain $g$, $r$, and $i-$band photometry. The ALFOSC data was reduced using the PyNOT\footnote{ https://github.com/jkrogager/PyNOT} pipeline. %We also obtained infrared images in the $J$ and $K$ bands with the Wide-field Infrared Camera \citep[WIRC;][]{Wilson2003}, on the Palomar 200-inch telescope. The WIRC images were reduced using a pipeline described in \citet{De2020c}. %We obtained one epoch of imaging in the $J$ band with the Multi-Object Spectrometer for Infra-Red Exploration  \citep[MOSFIRE;][]{McLean2010, McLean2012} instrument on the Keck-I telescope (\textcolor{black}{add reduction details -- ask Kishalay}).
The collage of all the multi-band lightcurves is shown in Figure \ref{fig:phot_all}. The photometry data is listed in Table \ref{tab:all_phot}. 

%\textcolor{black}{IR, LDT photometry yet to be subtracted -- ask Viraj and Igor. Also add any new photometry info WASP, LCO?, WINTER? LRIS?}

%\begin{figure*}
%    \centering
%    \includegraphics[width=13.5cm]{23zaw_spec_collage.png}
%    \caption{Spectral sequence for SN 2023zaw. The phases are in rest-frame days since the $r$-band peak (MJD = 60287.7). See Section \ref{section:spectra} for details on the obtained spectra.}
%    \label{fig:spec_all1}
%\end{figure*}

\subsection{Optical spectroscopy}
\label{section:spectra}
We acquired four epochs of spectroscopy with the Spectral Energy Distribution Machine \citep[SEDM;][]{Blagorodnova2018} on the Palomar 60-inch telescope. The SEDM data was reduced using the pipeline described in \citet{Rigault2019} and \citet{Kim2022}. We obtained three epochs of spectroscopy with the Low-Resolution Imaging Spectrometer \citep[LRIS;][]{Oke1995} on the Keck I telescope, with data reduced using the automated \lpipe{} \citep{Perley2019} pipeline. The Alhambra Faint Object Spectrograph and Camera (ALFOSC) instrument at the 2.5m Nordic Optical Telescope was used to obtain one epoch of low-resolution spectrum, which was reduced using the PyNOT\footnote{https://github.com/jkrogager/PyNOT} pipeline.  We took one spectrum with the Binospec spectrograph \citep{fabricant2019} on the Multiple Mirror Telescope, which was reduced using the \texttt{PypeIt} package \citep{pypeit:joss_pub, pypeit:zenodo}.
We used the Keck Cosmic Web Imager \citep[KCWI;][]{Martin2010, Morrissey2018} to obtain spectra of the transient and its host environment. This was reduced using the KCWI data reduction pipeline\footnote{https://kcwi-drp.readthedocs.io}. We also show one spectrum observed with the Gemini-GMOS instrument from the Transient Name Server\footnote{https://www.wis-tns.org/}. The collage of all the spectra is shown in Figure \ref{fig:spec_all} and the spectroscopy log can be found in Table \ref{tab:spec_log}.

%\subsection{Host galaxy photometry and spectroscopy}
%\textcolor{black}{To be added by Steve...}

\section{Methods and analysis}

\subsection{Extinction Correction}
\label{sec:extinction}
We correct for Milky Way extinction using the dust maps from \citet{Schlafly11}. Along the line of sight of SN 2023zaw, $E(B-V) = 0.25$ mag. For reddening corrections, we use the extinction law in \citet{Cardelli1989} with $R_V = 3.1$. 

For host extinction correction, we use the \ion{Na}{1} D absorption lines of the host-galaxy \citep{Poznanski2012, Stritzingetr2018}. 
We measure an equivalent width of $1.5 \pm 0.2\ \mathrm{\AA}$, from the KCWI spectrum obtained at a phase of $+2.5$ days past $r$-band peak. To compute $A_\mathrm{V}$, we use $A_\mathrm{V}^{\mathrm{host}} [\rm mag] = 0.78 (\pm 0.15) \times EW_{Na\ {\sc I}\ D} [\mathrm{\AA}]$ \citep{Stritzingetr2018}. We note that there are caveats in using the \ion{Na}{1} D EW measurement to estimate host extinction, especially for low-resolution spectra \citep{Poznanski2011} and in the presence of circumstellar material \citep{Phillips2013}.

We plot the $g-r$ and $r-i$ colors of SN 2023zaw before and after host extinction correction in Figure \ref{fig:color} and compare with other rapidly evolving transients and the SN Ib color template from \citet{Stritzingetr2018}.

\subsection{Explosion epoch estimation}\label{sec:explsoion}

We fit for the $r-$band maximum epoch with a polynomial fit to the $r-$band photometry and obtain a peak magnitude of $-16.7$ mag and peak MJD of 60287.7. All phases mentioned in the paper will be in rest frame days measured from this $r-$band peak epoch. The last non-detection before the first detection was on 60283 MJD (\textcolor{black}{-4.7 days}), with an upper limit of $-$14.8 mag in the $o-$band. We perform a power-law fit to the $r$-band data prior to the maximum epoch to estimate the explosion epoch. We find that the explosion epoch is 60284.4 $\pm$ 0.5 MJD (\textcolor{black}{$-3.3 \pm 0.5$ days}).

%We also fit for the explosion epoch as a free parameter in the shock-cooling model fit (see Section \ref{sec:scfit}). From both methods, we find that the explosion epoch is 60284.4 $\pm$ 0.5 MJD (\textcolor{black}{$-3.3 \pm 0.5$ days}).

\subsection{Lightcurve properties}

In Figures \ref{fig:phot_all} and \ref{fig:rbandcompare_r}, we compare the $g-$ and $r-$band evolution of SN 2023zaw to other fast-evolving and subluminous transients in the literature $-$ fast-declining Type I SNe $-$ SN 2005ek \citep{Drout2013}, SN 2010X \citep{Kasliwal2010}, SN 2018kzr \citep{McBrien2019}, SN 2019bkc \citep{Prentice2020, Chen2020}, the kilonova AT 2017gfo \citep{Abbott2017}, ultra-stripped SN candidates $-$ SN 2019dge \citep{Yao2020}, iPTF14gqr \citep{De2018c}, SN 2019wxt \citep{Shivkumar2023, Agudo2023}, SN 2022agco \citep{Yan2023}. We also compare with lightcurve templates for Type Ibn \citep{Hosseinzadeh2017} and Type Ibc SNe \citep{Taddia2015}. SN 2023zaw is much fainter and faster evolving than the typical Type Ibn or Ibc SN. \textcolor{black}{However, we note that some Type Ibn/Icn SNe also show rapid evolution: LSQ13ccw \citep{Hosseinzadeh2017}, SN~2023emq \citep{Pursiainen2023}}. We fit a polynomial to the lightcurves and measure the rise time and fade time as the duration above the half-maximum of the peak. SN 2023zaw has a rise time ($\mathrm{t^{1/2}_{rise}}$) of 1.8 days and a decline time ($\mathrm{t^{1/2}_{dec}}$)  of 3.1 days. The time above half maximum of the peak is shorter than for all other fast-evolving faint transients in the literature (see Figures \ref{fig:phot_all}, \ref{fig:timescale},  and \ref{fig:rbandcompare_r}). The lightcurve declines by 2.7 mag in 10 days in the $r-$band and by 2.7 mag in 6 days in the $g-$band. SN 2023zaw is only slower than kilonova AT 2017gfo and has a comparable decline timescale to SN 2019bkc and SN 2018kzr. In the $r-$band, SN 2023zaw has a very sharp initial decline of $\sim$ 0.28 mag $\mathrm{day}^{-1}$ followed by a transition to a relatively slower decline rate of $\sim$ 0.07 mag $\mathrm{day}^{-1}$ after $\sim$10 days since peak. In the $g$ band, SN 2023zaw shows a very sharp initial decline of $\sim$ 0.44 mag $\mathrm{day}^{-1}$ till 10 days post-peak. The peak absolute magnitude in the $r$ band before and after host-extinction correction is $-$15.6 mag and $-$16.7 mag respectively. In the $g-$band, the peak absolute magnitude before and after host-extinction correction is  $-15.2$ mag and $-16.6$ mag respectively. We compare the time above half maximum and the absolute luminosity of SN 2023zaw to a sample of SESNe detected by ZTF and other fast and faint SNe in the literature in Figure \ref{fig:timescale}. SN 2023zaw shows the fastest evolution among all transients at similar luminosities.

%peak, explosion epoch fit, etc

\begin{figure*}
    \centering
    \includegraphics[width=16.5cm]{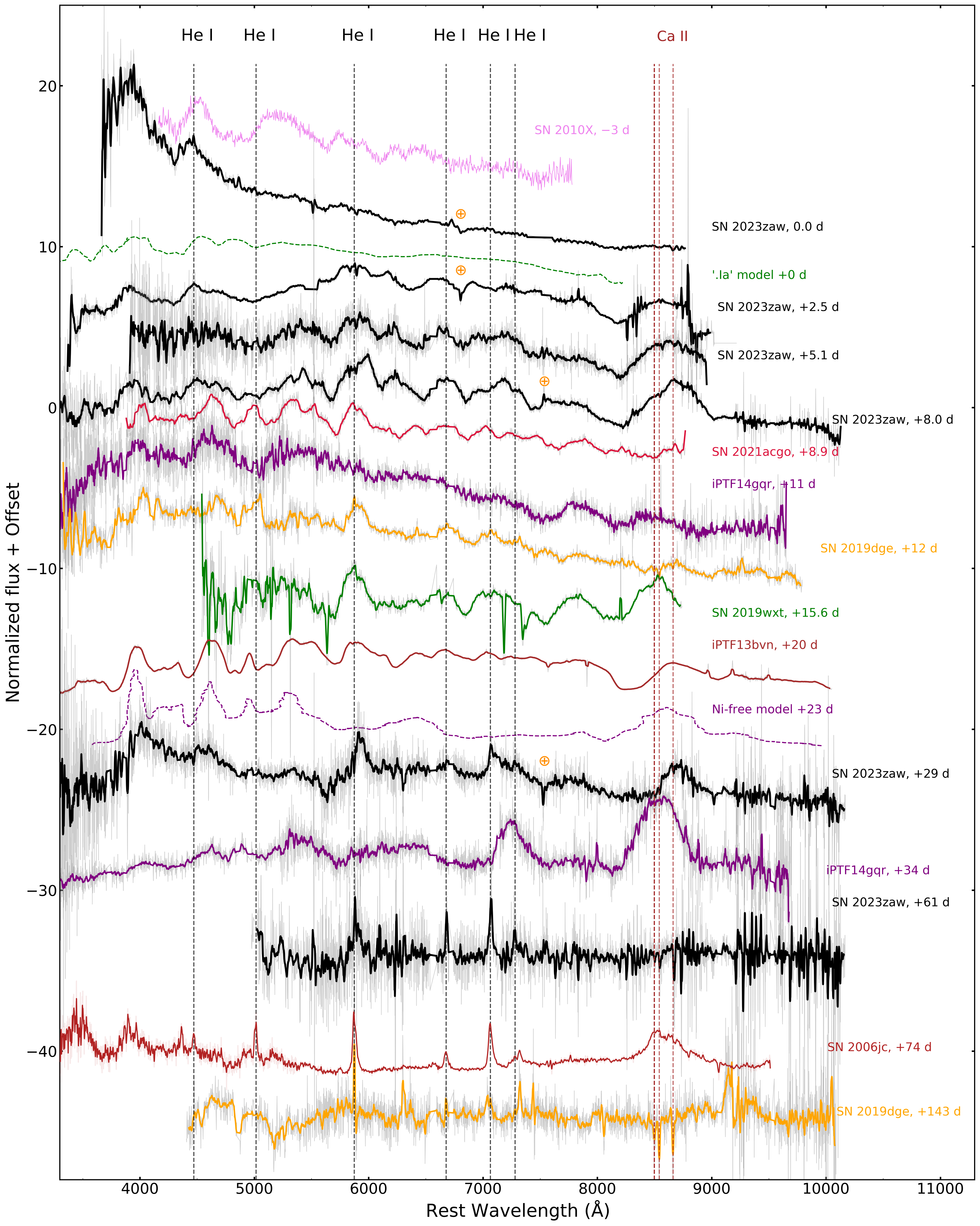}
    \caption{The spectra of SN 2023zaw (in black) compared with other fast-evolving transients in the literature $-$ fast-declining Type I SNe $-$ SN 2010X \citep{Kasliwal2010}, ultra-stripped SNe candidates $-$ SN 2019dge \citep{Yao2020}, iPTF14gqr \citep{De2018c}, SN 2019wxt \citep{Shivkumar2023, Agudo2023}, SN 2022agco \citep{Yan2023}, canonical Type Ib SN $-$ iPTF14bvn \citep{Fremling2018}, Type Ibn $-$ SN 2006jc \citep{Foley2007}. We also compare with theoretical predictions for `.Ia' SNe \citep{Shen2010} and nickel-free SN models \citep{Kleiser2014}. The spectra are corrected for Milky Way and host extinction. The $x$-axis represents the rest wavelength. The \ion{He}{1} and \ion{Ca}{2} emission lines are shown in dashed vertical lines. The telluric absorption features are indicated by the $\oplus$\ symbol. }
    \label{fig:spec_compare}
\end{figure*}

\subsection{Spectral Properties}

%We do not see any narrow flash ionization lines.

In the early SEDM spectra taken before $r-$band peak, we see a blue, featureless continuum. %However, we cannot rule out early flash ionization lines because of the low SNR and resolution of these spectra. 
We see multiple broad 
%He {\sc i} emission lines -- He {\sc i} $\lambda5876$, He {\sc i} $\lambda6678$ and He {\sc i} $\lambda7065$ 
helium lines -- \ion{He}{1} $\lambda\lambda 5876,6678,7065$,
in the subsequent spectra. %the ALFOSC spectrum taken at peak, the KCWI spectrum taken 2.5 days after peak, the GMOS spectrum \citep{Gillanders2023} taken 4 days after peak and the LRIS spectrum taken 8 days after peak. 
The expansion velocity measured from the absorption in the P-Cygni profile of the \ion{He}{1} lines evolved from 12,000 to 9700 $\mathrm{km\ s^{-1}}$ from 0 days to 8 days post-peak. These \ion{He}{1} absorption lines resemble those seen in Type Ib SNe and we obtain a decent fit to Type Ib SNe at similar or later phases using the \texttt{SuperNova Identification} \citep[\texttt{SNID};][]{Blondin2007} code. This is consistent with the initial classification made by \citet{Gillanders2023}. There is also a redshifted component to the \ion{He}{1} $\lambda5876$ line profile \textcolor{black}{in the spectrum taken at $+8$ and $+29$ days}, probably due to asymmetrical ejecta. \ion{He}{1} line velocities drop to $\sim3000$  $\mathrm{km\ s^{-1}}$ at a phase of 29 days. Prominent narrow \ion{He}{1} emission lines can be seen in the LRIS spectra obtained at $29$ and $61$ days after peak. The FWHM velocities of the \ion{He}{1} emission lines in the latest LRIS spectrum are $\sim 1050 \pm 100$ $\mathrm{km\ s^{-1}}$.

In the photospheric phase, we do not see any broad \ion{O}{1} $\lambda7774$ emission line, which is common in stripped-envelope SNe. We also see broad, high-velocity lines of the \ion{Ca}{2} NIR triplet at $\sim 12,000~\mathrm{km\ s^{-1}}$ as early as 2 days after peak. Such high-velocity early Ca NIR absorption lines are not seen in other fast-evolving SNe (see Figure \ref{fig:spec_compare}). Also, unlike other rapid-declining SNe, SN 2023zaw does not reach nebular phase even at 29 days after peak. Broad nebular lines such as [O~I] $\lambda \lambda$6300, 6364, \ion{Ca}{2} NIR triplet are also not seen in the latest LRIS spectrum taken at +61 days. %The absence of broad O {\sc i} $\lambda$7774 and the presence of early Ca {\sc ii} NIR lines resemble SN 2010X \citep{Kasliwal2010} which is a ``.Ia'' or Ni-free SN candidate. %However, the presence of He {\sc i} was not conclusive for SN 2020X. 

%\textcolor{black}{any other spectral feature that should be mentioned?, mention other metal line absorptions?, comment on presence/absence of intermediate mass elements?}

%\begin{figure*}
%    \centering
%    \includegraphics[width=6.5cm]%{23zaw_bol_SN.png}\includegraphics[width=6.5cm]%{23zaw_radius_SN.png}\includegraphics[width=6.5cm]%{23zaw_temp_SN.png}
%    \caption{}
%    \label{fig:BBcompare}
%\end{figure*}

\subsection{Bolometric Lightcurve}
\label{sec:blackbody}

We bin the data in intervals of one day and fit with the Planck blackbody function where there are detections in at least three filters. We perform the fit using the Python \emcee\ package \citep{Foreman-Mackey13}. The blackbody temperature, radius and luminosity thus obtained are shown in Table \ref{table:bb_table}. The errors in the model parameters are determined by extracting the $16^\mathrm{th}$ and $84^\mathrm{th}$ percentiles of the posterior probability distribution. For the later epochs, where we have photometry in only one band, we use the extrapolated blackbody temperatures \textcolor{black}{obtained by fitting an exponential decay function to the temperature data from previous epochs}. The temperature stays roughly constant at $\approx$ 4000 K. For these later epochs, we also estimate the bolometric luminosity as $\nu L_\nu$. These estimates agree with each other within 3$\sigma$, with the latter method slightly underestimating the bolometric luminosity as expected.

\begin{figure*}
    \centering
    \includegraphics[width=6.5cm]{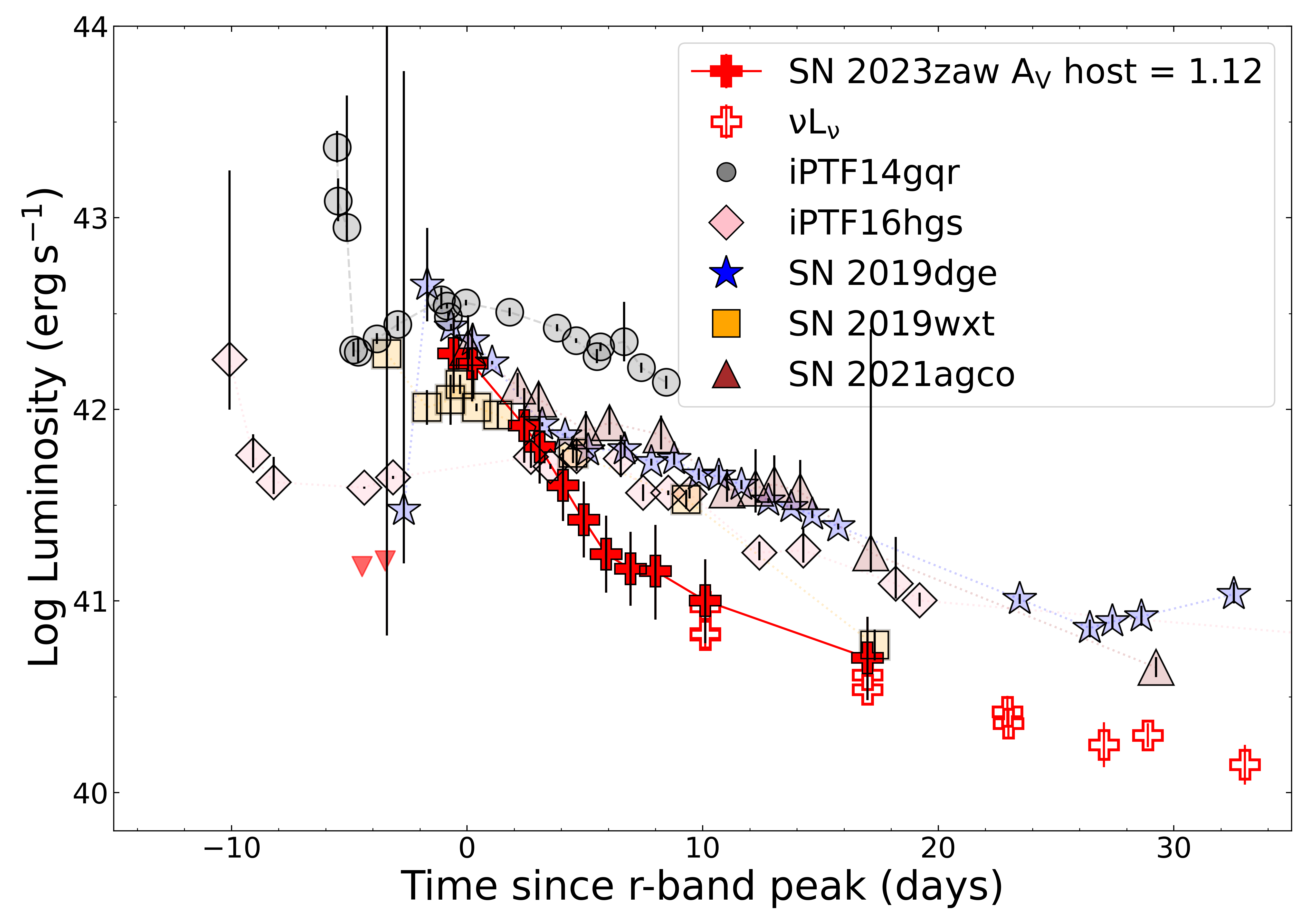}\includegraphics[width=6.5cm]{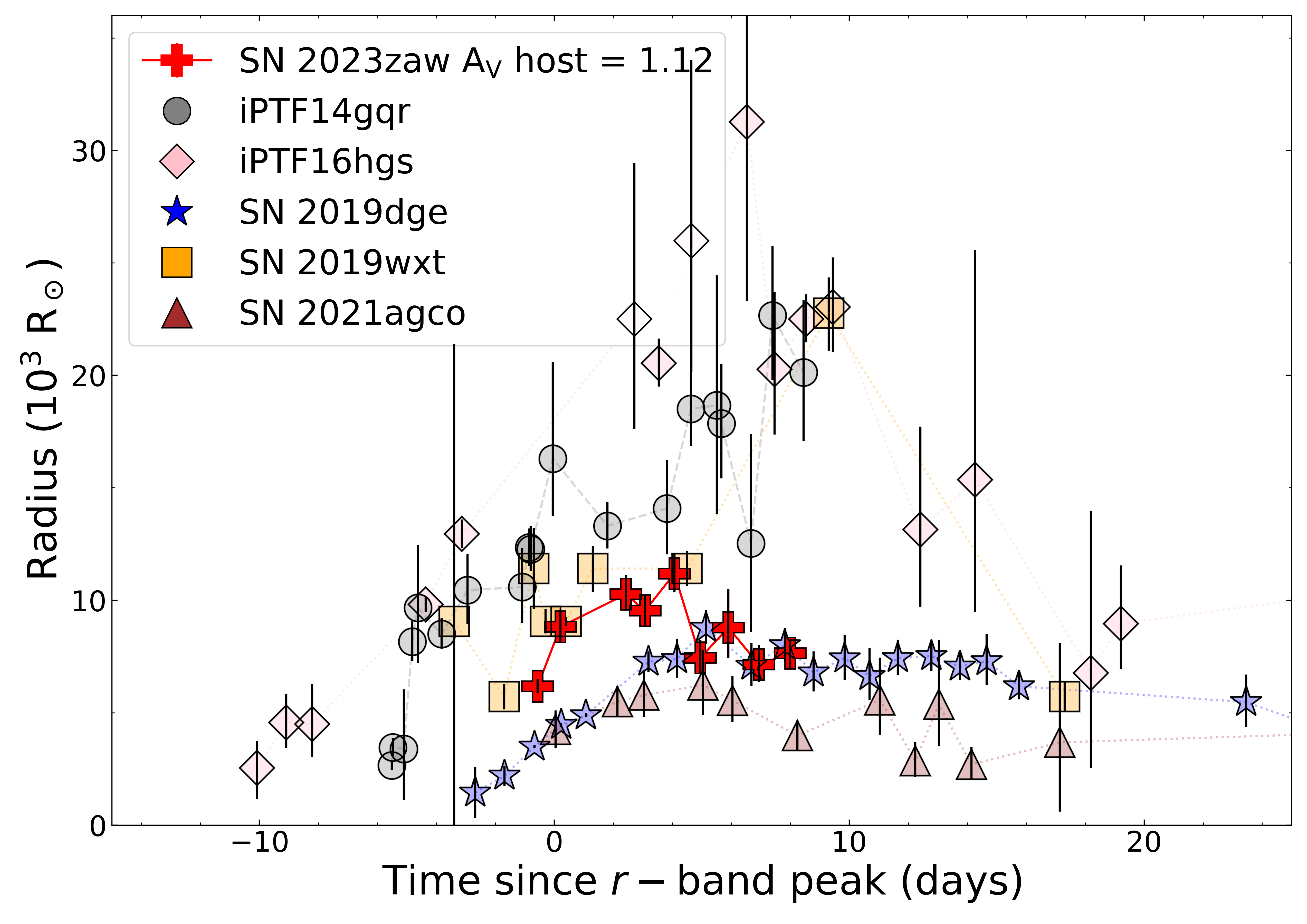}\includegraphics[width=6.5cm]{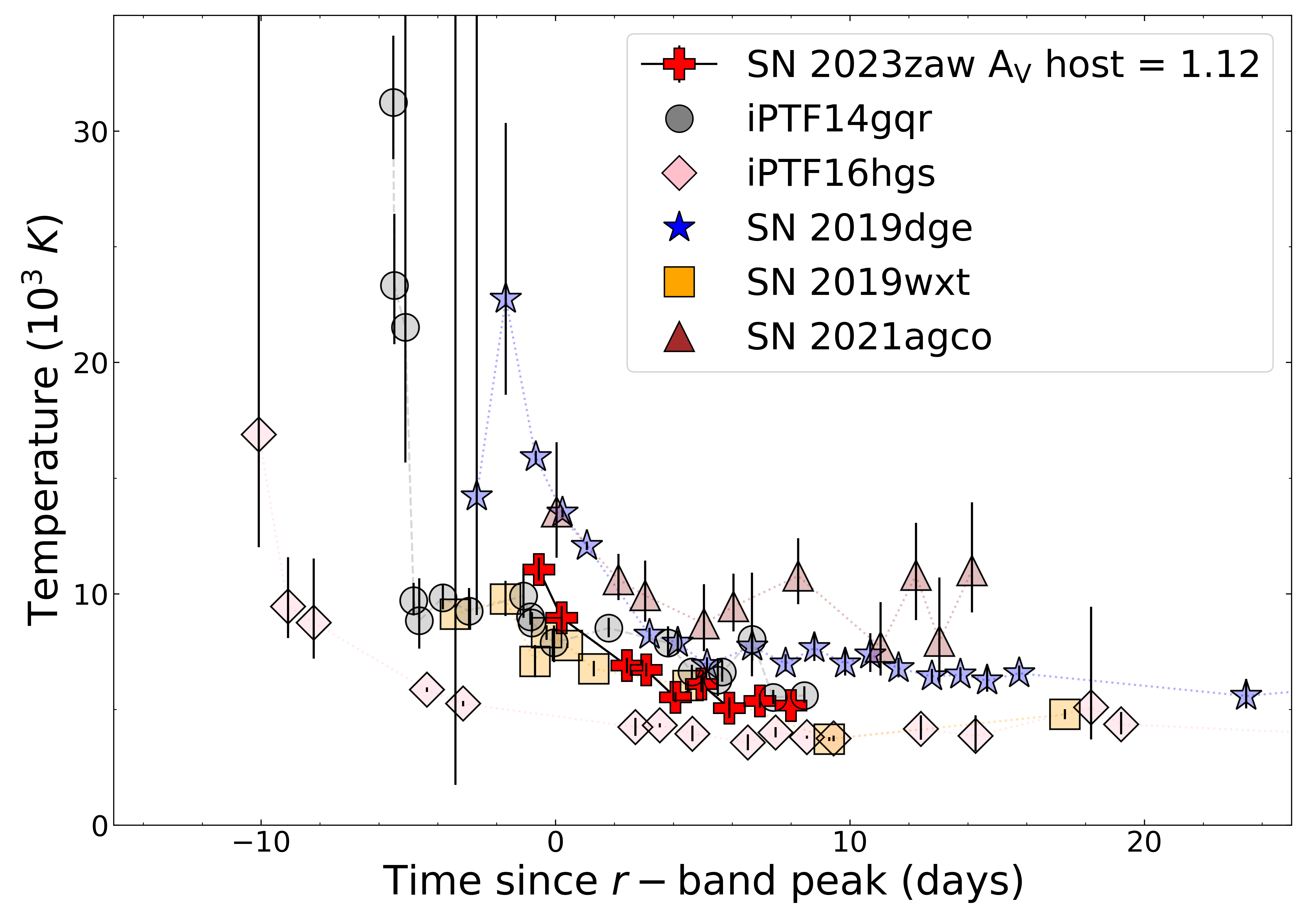} \\
        \includegraphics[width=8.5cm]{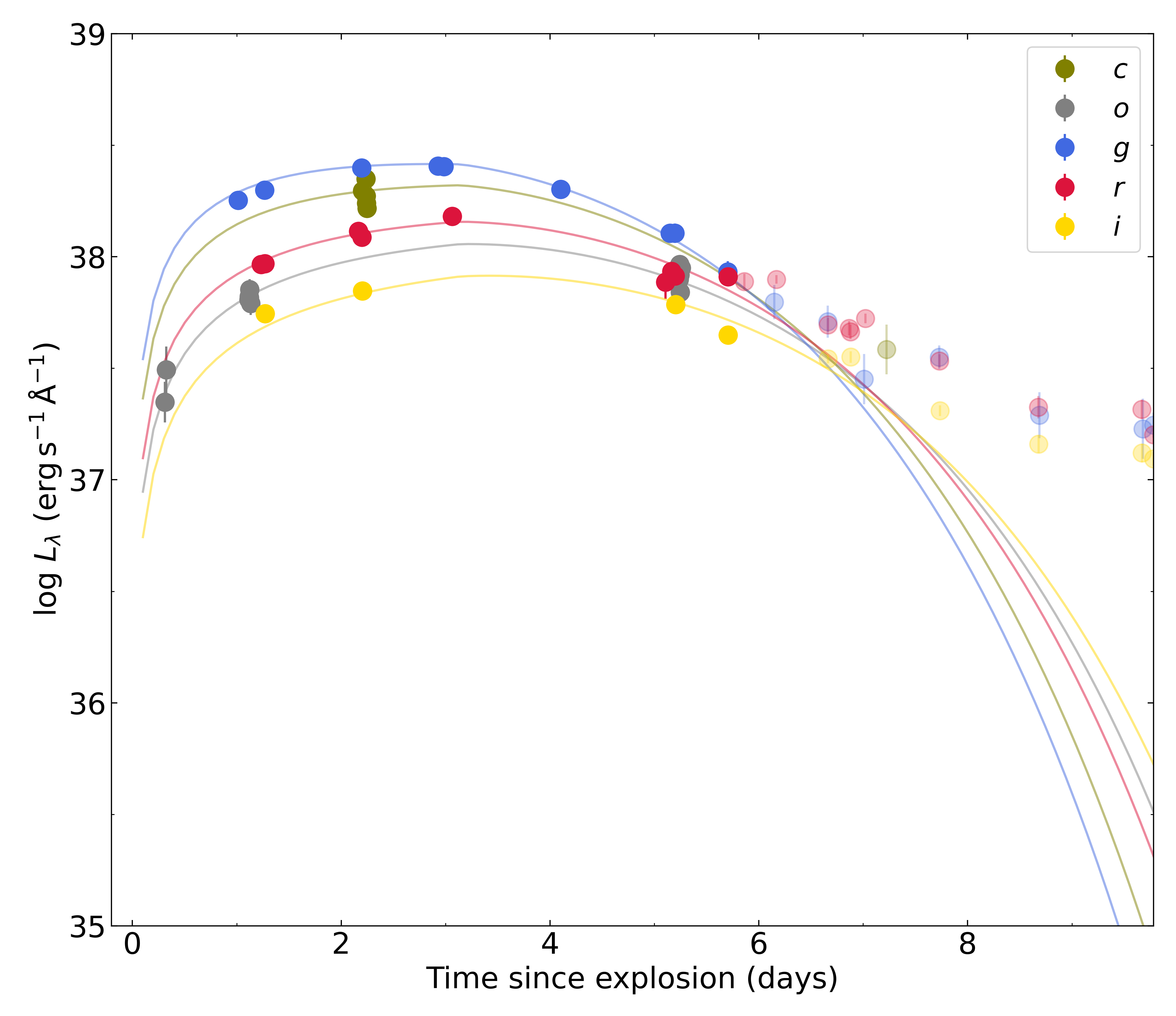}\includegraphics[width=8.5cm]{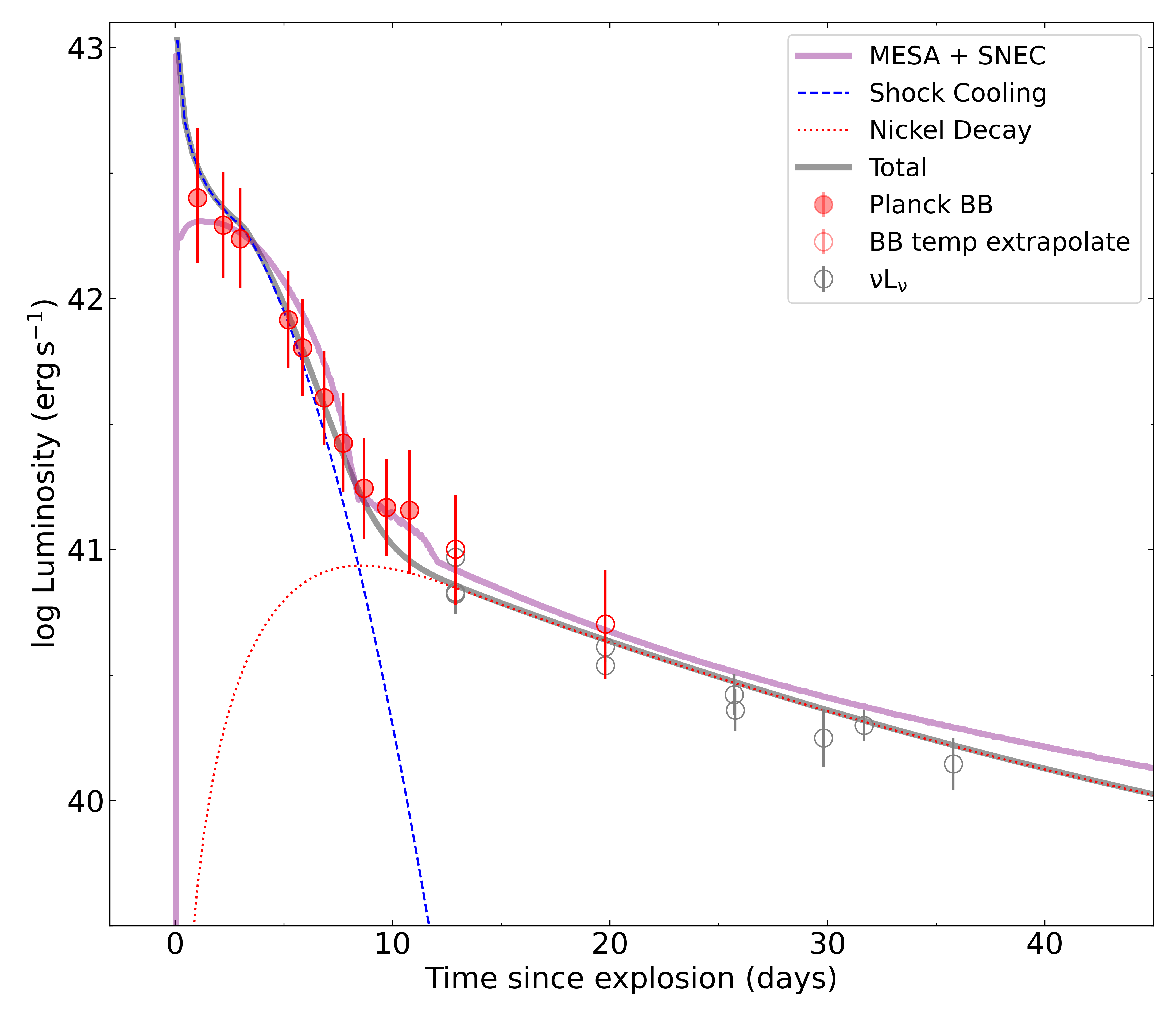} 
    
    \caption{\textit{Top:} Comparison of the bolometric luminosity, best-fit radius and temperature evolution of SN 2023zaw from the  blackbody fits described in Section \ref{sec:blackbody} with ultra-stripped SN candidates $-$ SN 2019dge \citep{Yao2020}, iPTF14gqr \citep{De2018c}, SN 2019wxt \citep{Shivkumar2023, Agudo2023}, SN 2022agco \citep{Yan2023}.  
    \textit{Bottom left:} Shock cooling emission model fit to the early-phase multiband lightcurve as described in Section \ref{sec:scfit}. \textit{Bottom right:}  Best-fits for bolometric luminosity with radiation-hydrodymanics and analytical models as described in Section \ref{sec:lightcurveanalysis}.} 
    \label{fig:shockfit}
\end{figure*}

We plot the best-fit blackbody parameters and compare these with other ultra-stripped SN candidates in Figure~\ref{fig:shockfit}. We can see that the bolometric light curve of SN 2023zaw is similar in peak brightness to the other SNe but has a faster decline rate within the first 10 days after $r$-band peak. The decline of the bolometric luminosity slows down after around 6 days post-peak. The photospheric radius increases up to 4 days post-peak, reaching a maximum of $\sim$10 000 \Rsun. The temperature decreases over time, from a maximum of $\sim$17 000 K on day 0 to $\sim$4000 K on day 8. 

%\textcolor{black}{some non-physical fluctuations in radius, temp estimates?}

\subsection{Modeling the lightcurve}\label{sec:lightcurveanalysis}
\subsubsection{Radiation-Hydrodynamics Modeling}\label{sec:radhydro}
%\subsubsection{Pre-supernova models}\label{sec:presupernova}
We use the pre-supernova model grid presented in \citet{Wu2022b}. This grid was produced through binary evolution of He-stars with an initial mass of 2.5 \Msun $-$ 3 \Msun\, evolved up to silicon burning using the Modules for Experiments in Stellar Astrophysics code \citep[\textsc{mesa};][]{Paxton2010}. The binary systems evolved in \citet{Wu2022b} include orbital periods of 1, 10, and 100 days. Predictions of the properties of the unbound CSM in the vicinity of each pre-supernova model were also produced. In this work we compare our data to both models with and without unbound CSM.

%\subsubsection{Modeling the bolometric lightcurve}
We use the SuperNova Explosion code \citep[\textsc{snec};][]{Morozova2015} to explode the grid of pre-supernova stellar models described above. First, we exploded all models in the grid using a thermal bomb with $\mathrm{E}_\mathrm{kin}=0.4\times10^{51}$~erg, and $\mathrm{M}_{\mathrm{Ni}}=0.0035$~\Msun, with \Nif\ mixed all the way through the remaining star after mass excision of $1.4$~\Msun\ in the center.

The resulting bolometric lightcurves and expansion velocities were then compared to our data on SN~2023zaw. We find that the model \citet{Wu2022b} produced from a He-star with an initial mass of $2.651$~\Msun\ and an orbit period of 100 days reproduces the overall behavior seen in our data well, including the expansion velocities. Other models fail to reproduce both the early cooling-phase of the bolometric lightcurve (governed by the radius and mass in the bound envelope) or the late-time decline rate (governed by the total ejecta mass). To produce our best-fitting MESA$+$SNEC model shown in Figure~\ref{fig:shockfit}, we modified the bound envelope radius of the raw MESA model so that it extends to $\approx50$~\Rsun, by cutting the model grid appropriately (compared to $\approx110~\Rsun$ in the raw model). This would correspond to an orbital period between 10 and 100 days. We also modified the mass excision in the center to be $1.3$~\Msun. These modifications result in the explosion parameters $\mathrm{E}_\mathrm{kin}=0.4\times10^{51}$~erg, $\mathrm{M}_{\mathrm{Ni}}=0.0035$~\Msun, with complete mixing of \Nif\ through the ejecta, which has a total mass of $\mathrm{M}_{\mathrm{ej}}=0.56$~\Msun, including the bound envelope.
%The measured envelope properties are consistent with the bound and unbound CSM properties of low-mass He stars (see Figure \ref{fig:bound}). 
%From our radiation-hydrodynamic modeling with \snec, we find that the bound envelope extending to $\approx$ 50 \Rsun\ for a He-star with initial mass 2.6 \Msun\ provides a good fit to the observed lightcurve. Other models fail to reproduce either the early cooling-phase of the bolometric lightcurve (governed by the radius and mass in the bound envelope) or the late-time decline rate (governed by the total ejecta mass). We note that the models including the unbound CSM predicted by \citet{Wu2022b} are largely inconsistent with our data. Luminosities and durations can be similar to what we are observing if the explosion energy is adjusted to be low ($<0.1\times10^{51}$~erg). However, they show far too low expansion velocities during the first week after explosion compared to what we derive from our blackbody fitting and spectroscopic absorption line measurements.
Models including the unbound CSM predicted by \citet{Wu2022b} are largely inconsistent with our data. The luminosity during the first week can be similar to what we observe if the explosion energy is adjusted to be low ($<0.1\times10^{51}$~erg). However, such models are inconsistent (too bright) at later times and inconsistent with the overall decay rate.% However, they show far too low expansion velocities during the first week after explosion compared to what we derive from our blackbody fitting and spectroscopic absorption line measurements. %opacity_floor_envelope = 0.03d0 opacity_floor_core     = 0.010d0 \textsc{snec} is an open-source radiation$–$hydrodynamics Lagrangian code, useful for simulating stellar explosions and their lightcurves. \textcolor{black}{Text to be added by Christoffer; also add in discussion on bound, CSM, progenitor system, binary separation, etc. }

\subsubsection{Analytical Modeling}
\label{sec:scfit}

%\subsubsection{Fitting for shock-cooling}

The bolometric luminosity declines at a rapid rate for the first $\sim6$ days after peak (see Figure \ref{fig:shockfit}). %--add on why fit with SC?
Radioactivity alone is unable to power this early-time luminosity under the constraint that the ejecta mass must be greater than the nickel mass. The peak luminosity of $2.5 \times 10^{42}\ \mathrm{erg\ s^{-1}}$ and timescale of evolution ($\approx$ 4.9 days) falls in the forbidden region of nickel-powered lightcurves \citep[see Figure 1 in][]{Kasen2017}. Instead, we model this early lightcurve as being powered by shock-cooling emission from the shock-heated bound or unbound extended stellar material. We use the shock-cooling emission model presented in \citet{Piro2021} to fit the early-time multiband photometry. We obtain a good fit for all the filters for photometry up to 6 days past peak (Figure \ref{fig:shockfit}). The explosion time is constrained to be $-2.7 \pm 0.2$ days before the peak. Based on the fit, we find that the best-fit envelope mass is $0.20 \pm 0.01$ \Msun\ and envelope radius is $55 \pm 11$ \Rsun. This is comparable to the envelope properties measured from shock-cooling emission modeling in other Type Ibc SNe (see Figure \ref{fig:envelope}). The envelope properties of SN 2023zaw are also consistent with theoretical models for bound and unbound stellar material (Figure \ref{fig:bound}).

%N2019dge \citep{Yao2020}, iPTF14gqr \citep{De2018c}, SN 2019wxt \citep{Shivkumar2023, Agudo2023}, SN 2022agco \citep{Yan2023} and Ca-rich Type IIb SNe \citep{Das2023a}. \textcolor{black}{-- update figure}. The shocked material which gives rise to the early lightcurve is either the puffed up bound stellar material or unbound stellar material that is lost during the later stages of single and binary stellar evolution. We compare the predictions of various stellar evolution and mass-loss models in Figures \ref{fig:bound}. We find that the measured envelope properties are consistent with both the scenarios. 

\begin{figure*}
    \centering
    \includegraphics[width=13.5cm]{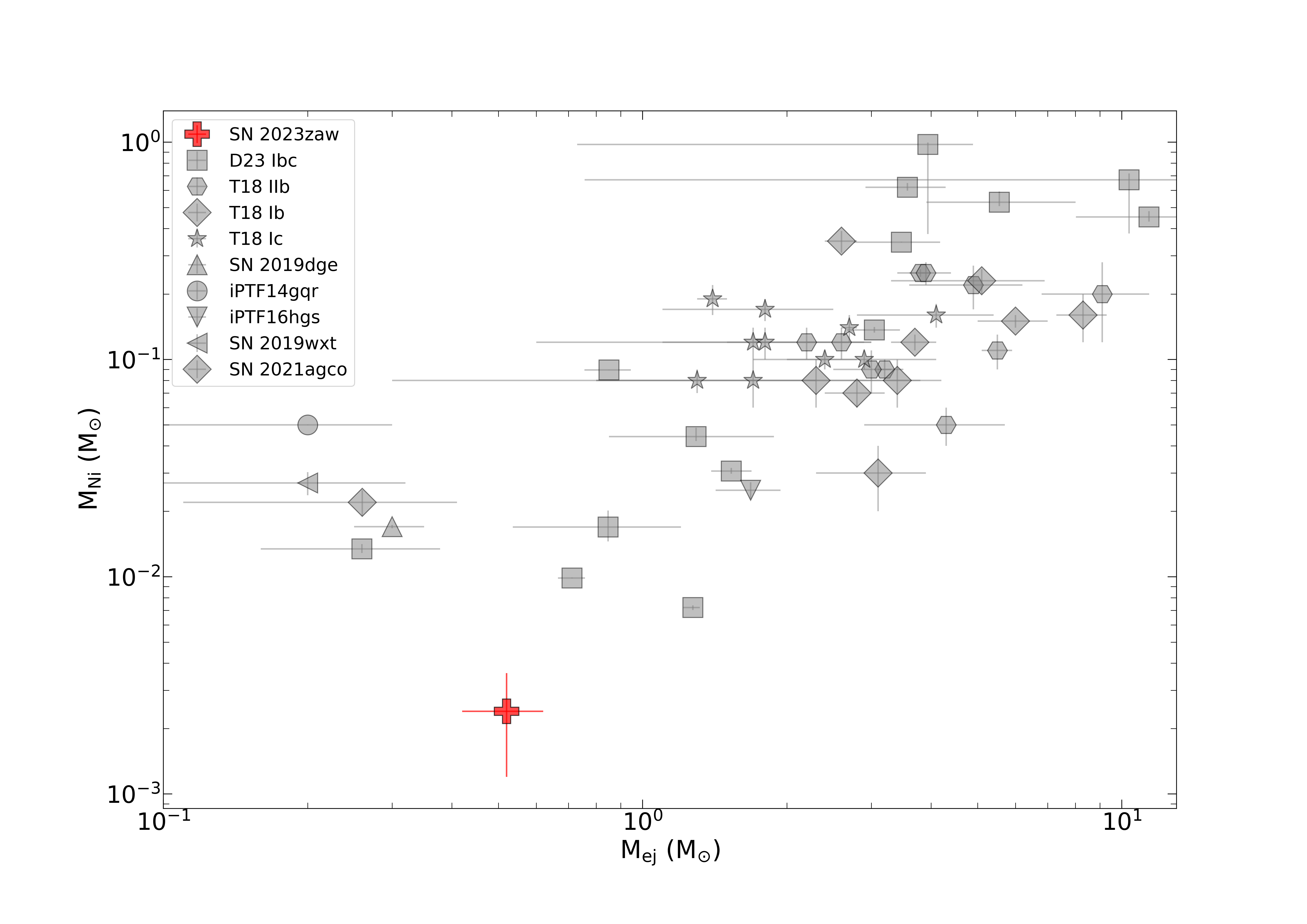}
    \caption{We compare the estimated ejecta mass and nickel mass with Type Ibc SNe from \citet{Taddia2018}, double-peaked Type Ibc SNe from \citet{Das2023b} and other ultra-stripped SN candidates $-$ SN 2019dge \citep{Yao2020}, iPTF14gqr \citep{De2018c}, SN 2019wxt \citep{Shivkumar2023, Agudo2023}, SN 2022agco \citep{Yan2023}. The red plus sign denotes the best-fit parameters by fitting shock cooling emission $+$ nickel decay to the lightcurve after host-extinction correction.}
    \label{fig:mnimej}
\end{figure*}

%\subsubsection{Fitting for Ni decay}

For estimating the Ni mass, we first subtract the contribution of the shock cooling emission estimated above from the bolometric luminosity, which we then assume to be entirely powered by radioactive decay. 
We use the analytical expression for a Ni-powered bolometric lightcurve provided in \citet{Arnett89}, \citet{Valenti2008}, and \citet{Wheeler2015}. Further details on the model fitting can be found in \citet[][their Appendix B]{Yao2020}. 
%(their Appendix B). 
Based on the fit, we find that the required \textcolor{black}{Ni-mass to power the supernova is $0.24 \pm 0.14 \times 10^{-2}$ \Msun}. We measure a photon diffusion timescale ($\tau_m$) of $6.21^{+0.11}_{-0.11}$ days and characteristic $\gamma$-ray diffusion timescale ($t_o$) of $36.04^{+3.41}_{-2.74}$ days. 

We assume that the photospheric velocity is $\sim12 000$ km~s$^{-1}$, as assumed from the \ion{He}{1} absorption lines in the ALFOSC spectrum taken at $r-$band peak. We use an opacity ($\kappa$) value of 0.07 $\mathrm{cm^2\ g^{-1}}$, appropriate for H-poor SNe \citep{Taddia2018}. We estimate the ejecta mass using the following relation: $\tau_m = \big(\frac{2\kappa \mathrm{M_{ej}}}{13.8c\mathrm{v_{phot}}}\big)^{1/2}$, which gives $\mathrm{M_{ej}}$ $\approx 0.52^{+0.02}_{-0.02}\ \times $ $\big(\frac{\mathrm{v_{phot}}}{\textcolor{black}{12000}\ 
\mathrm{km/s}}\big)$ \Msun. The calculated explosion kinetic energy is $\approx 4.82^{+0.18}_{-0.18}\times 10^{50}\ \mathrm{erg\ s^{-1}} $.

These explosion parameters are an order of magnitude lower than for regular Type Ibc SNe. The estimated ejecta mass is comparable to other ultra-stripped SNe in the literature while the Ni-mass is smaller by a factor of $\sim$5. The rapid light curve decline is explained by the low ejecta mass, which allows $\gamma$-rays to escape already a few days after the peak. Overall, SN 2023zaw occupies a unique location in the phase space plot of nickel mass versus ejecta mass (see Figure \ref{fig:mnimej}). For comparison, we also plot the nickel and ejecta masses estimated for a sample of stripped-envelope SNe from \citet{Taddia2018} and double-peaked Type Ibc SNe from \cite{Das2023b}. We note that the double-peaked SN 2021inl \citep{Jacobson2022b, Das2023b} is the other SN with Ni-mass $<$ 0.01 \Msun.

We also use the analytical models given in \citet{KhatamiKasen2019} to obtain a nickel mass estimate of $\approx2.5 \times 10^{-2}$ \Msun\ and an ejecta mass estimate of $\approx 0.45$ \Msun.

\begin{table*} 
\begin{center} 

%\begin{table*}[htbp]
\caption{Comparison of explosion properties of SN 2023zaw with ultra-stripped SN candidates $-$ SN 2019dge \citep{Yao2020}, iPTF14gqr \citep{De2018c}, iPTF16hgs \citep{De2018d}, SN 2019wxt \citep{Shivkumar2023, Agudo2023}, SN 2022agco \citep{Yan2023}.}
    \centering
    
    % \begin{threeparttable}
        \resizebox{\linewidth}{!}{
        \begin{tabular}{ccccccccc}
        % \begin{tabular}{cccccccc}
            \hline
            Source & Redshift & Host Type &  $r-$band Peak & $M_{\rm ej}$            & $M_{\rm Ni}$                 & $R_{\rm ext}$ & $M_{\rm ext}$ \\
                  &          &         &  (mag)       &  (\msun) & ($10^{-2}$~\msun) &  ($10^{13}$~cm) &  ($10^{-2}$~\msun)\\
            
            \hline           
            SN 2023zaw& 0.0101& Spiral  & $-16.7 \pm 0.1$ & \textcolor{black}{$0.52^{+0.02}_{-0.02}$} & $0.24_{-0.02}^{+0.02}$ & $0.4_{-0.1}^{+0.1}$& $19.53_{-0.98}^{+0.99}$   \\
            
           SN 2021agco& 0.01056& Spiral  & $-16.2 \pm 0.24$ & $0.26_{-0.02}^{+0.04}$ & $2.2_{-0.3}^{+0.2}$ & $0.55_{-0.14}^{+0.18}$& $10.04_{-1.05}^{+1.87}$   \\
            % $0.23_{-0.06}^{+0.04}$ & $2.0_{-0.60}^{+0.90}$        & $0.7^{+0.28}_{-0.14}$  & $10.0_{-2.0}^{+1.0}$ \\
            
            AT 2019wxt & 0.036  & Compact &   $-16.6 \pm 0.4$ & $0.20^{+0.12}_{-0.11}$ & $2.7^{+0.33}_{-0.18}$ & $35.8^{+4.06}_{-3.68}$ & $3.55^{+0.12}_{-0.11}$\\
            
            SN 2019dge & 0.021  & Compact &  $-16.3 \pm 0.2$ & $0.30^{+0.02}_{-0.02}$ & $1.6^{+0.04}_{-0.03}$ & $1.2^{+0.06}_{-0.05}$  & $9.71^{+0.28}_{-0.27}$\\
            
            iPTF14gqr & 0.063  & Spiral  &  $-17.5 \pm 0.2$ & $0.20^{-0.10}_{+0.10}$ & $5.0^{+0.14}_{-0.15}$   & $6.1^{+8.73}_{-3.18}$  & $2.59^{+0.46}_{-0.34}$\\
            
            iPTF16hgs & 0.017  & Spiral  &  $-15.5 \pm 0.2$ & $1.68^{+0.28}_{-0.25}$ & $2.5^{+0.20}_{-0.22}$ & $2.6^{+14.08}_{-1.80}$ & $9.27^{+3.40}_{-2.48}$\\
            \hline
        \end{tabular}\label{table:ultracompare}
        }
        % $0.10_{-0.01}^{+0.02}$ & $78.37_{-19.94}^{+25.59}$ & $8.93_{-1.61}^{+2.59}$ & $9.57_{-1.62}^{+3.01}$ & $0.26_{-0.02}^{+0.04}$ & $0.022_{-0.003}^{+0.002}$ $^a${} 
        % respectively.}
    % \end{threeparttable}
%\end{table*}
\end{center} 
\end{table*}

\subsection{Host galaxy environment} \label{sec:host}

%\textcolor{black}{To be added by Steve...}
We observed SN\,2023zaw and its immediate environment with the integral-field unit KCWI (Section \ref{section:spectra}). %Spectra of various regions show continuum emission from evolved stellar populations and super-imposed narrow emission lines produced by the ionised gas of star-forming regions. 
These data allow us to address two outstanding questions: i) did SN 2023zaw explode in a star-forming region, and ii) how high is the host attenuation along the line of sight towards SN 2023zaw? 

We cleanly detect emission lines from a star-forming region at the SN position in the unbinned KCWI data. To measure the properties of the star-forming region, we extract the SN spectrum using a $4\times4$-px aperture, tie the flux scale to our $g$- and $i$-band photometry, and extract the flux of H$\alpha$-H$\gamma$, [\ion{O}{2}]$\lambda\lambda$3729,3729, [\ion{O}{3}]$\lambda\lambda$4959,5007, [\ion{N}{2}]$\lambda\lambda$6548,6584, and [\ion{S}{2}]$\lambda\lambda$6717,6731 by fitting each line with a Gaussian. All measurements are summarised in Table \ref{tab:explosionsite}.

We infer the attenuation by ionised gas from the flux ratio between H$\beta$ and H$\gamma$ \citep{Momcheva2013a}. %, which both fall on the blue arm of KCWI. 
Using the \citet{Calzetti2000a} attenuation model, we measure $E(B-V)_{\rm gas} = 0.58 \pm 0.20$~mag, which translates to a stellar continuum colour excess of $0.26 \pm 0.09$~mag. This value is comparable to the reddening inferred from the Na I D absorption lines (Section \ref{sec:extinction}) and corroborates that the line of sight is suffering from significant host attenuation.

We note that the stellar populations near SN\,2023zaw are evolved, and their spectra show conspicuous Balmer absorption lines. This could introduce a bias into the observed Balmer decrement. To quantify the impact, we centre a box annulus on the SN position and extract a spectrum from the blue KCWI data between 2 to 4~px from the SN position. We fit the stellar continuum with the stellar population fitting code \textsc{firefly} version 1.0.3 \citep{Wilkinson2017a, Neumann2022a}, utilising the stellar population models from \citet{Maraston2020a} and assuming the Kroupa initial mass function (IMF), a spectral resolution of 1000--1250 appropriate for our spectroscopic observation \citep{Morrissey2018a} and a velocity dispersion of $130\pm20~\rm km\,s^{-1}$. The best-fit yields an  $E(B-V)_{\rm star}$ of 0.36~mag, comparable to the value at the explosion site. Henceforth, we use $E(B-V)_{\rm gas} = 0.58 \pm 0.20$~mag to characterise the properties of the star-forming region further.  %After subtracting the stellar continuum model, the decrement between  H$\beta$ and H$\gamma$ gives a gas-phase attenuation of $\sim0.18$~mag. While this value is consistent with the value from the explosion site, it is highly sensitive to the chosen spectral resolving power.

The attenuation-corrected H$\alpha$ flux of $1.2\times10^{-14}~\rm erg\,cm^{-2}\,s^{-1}$ translates to a star-formation rate of $1.4^{+1.2}_{-0.6}\times10^{-2}~M_\odot\,\rm yr^{-1}$ utilising the relationship between the H$\alpha$ luminosity and the star-formation rate from \citet{Kennicutt1998a} and the scaling factor from \citet{Madau2014a} to convert from the Salpeter to the Chabrier IMF in the \citet{Kennicutt1998a} relation. The error estimate includes the uncertainty of the flux measurements, the gas-phase attenuation and the luminosity distance of SN\,2023zaw's host galaxy. We stress that the large SFR error reflects the uncertainty in the attenuation correction, not a low precision of the H$\alpha$ flux measurement (Table \ref{tab:explosionsite}). 

Detecting emission lines from hydrogen, nitrogen and oxygen also allows us to measure the metallicity of the star-forming region. Using the O3N2 metallicity indicator and the calibration from  \citet{Curti2017a}, we infer a metallicity of 1.18 $\pm$ 0.02 solar (statistical error), adopting a solar oxygen abundance of 8.67 \citep{Asplund2009a}.

Next, we draw our attention to SN\,2023zaw's neighbourhood. Figure \ref{fig:galaxy} shows a map of the star-formation activity out to a distance of 6.7~kpc from SN\,2023zaw's location. This image was generated from the KCWI data. First, we extracted a 20-\AA-wide image centred on the wavelength of H$\alpha$. Then, we extracted two additional 20-\AA-wide images from emission-line free regions adjacent to H$\alpha$ and interpolated between them to remove the galaxy/SN flux at the wavelength of H$\alpha$. Finally, the intensity scale was corrected for MW extinction, scaled to the SN photometry and converted to star-formation rate as we did for the explosion site. The SFRs in Figure \ref{fig:galaxy} are strict lower limits. They account for neither host attenuation nor stellar Balmer absorption lines.

The explosion site, marked by the blue circle in Figure \ref{fig:galaxy}, is located near the outskirts of a larger region with ongoing star-formation activity. Although the level of star formation activity at the explosion site is in the lower half of the intensity distribution, this star-forming region extends much farther. Inspecting the neighbouring spaxels reveals that the star-formation activity extends to regions of even less intensity. %This provides circumstantial evidence that SN\,2023zaw is indeed the product of a massive star explosion. 
\textcolor{black}{Thus, the location of SN\,2023zaw favors a massive star progenitor, but it does not rule out a thermonuclear origin.}

%Thus, the explosion site strongly favors a massive star progenitor.

%\begin{figure*}
%    \centering
%    \includegraphics[width=6cm]{23zaw_galaxy.png} \includegraphics[width=8.5cm]{SN2023zaw_KCWI_Halpha.pdf}
%    \caption{\textit{Left}: Pan-STARRS image UGC 03048, the host-galaxy of SN 2023zaw. The location of the transient is indicated by the white cross at the center. \textit{Right}: \textcolor{black}{In progress...H-alpha map of the host environment from the KCWI spectrum. Placeholder for similar plot}}
%    \label{fig:host}
%\end{figure*}

\section{Discussion: Progenitor and evolutionary pathway}

%\subsection{Progenitor and evolutionary pathway}

In this section, we discuss the possible progenitors and evolutionary pathways for SN 2023zaw. 

%\subsection{Is there pre-SN mass loss?}

%The shocked material which gives rise to the early lightcurve is either the puffed up bound stellar material or unbound stellar material that is lost during the later stages of single and binary stellar evolution. We compare the predictions of various stellar evolution and mass-loss models in Figures \ref{fig:bound}. We find that the measured envelope properties are consistent with both the scenarios. 

%\subsection{Others?}

%discuss these at the start and dismiss

%Partial explosions of C/O white dwarfs near the Chandrasekhar mass (Kromer 2013)

%Accretion-induced collapse (AIC) of a white dwarf to a neutron star See Kleiser 2014 for references -- needs 0 Ni mass?

%Kilonova?

\subsection{Thermonuclear `.Ia' SN}

In the `.Ia' SN scenario, the progenitor is a white dwarf in a close binary (P$_{\rm{orb}}<1$ hour) system, accreting helium from its companion. They have a tenth of the ejecta mass and explosion energy of a normal Type Ia supernova \citep{Bildsten2007}. We compare the $r$-band lightcurve and bolometric luminosity with `.Ia' SN models from \citet{Shen2010} in Figure \ref{fig:modelcompare}. The peak luminosities range from $0.5 - 5 \times 10^{42}\ \mathrm{erg~s^{-1}}$ or peak bolometric magnitudes from $-15.5$ to $-18$ mag, consistent with SN 2023zaw. We highlight the best-fit `.Ia' SN bolometric lightcurve which is an excellent fit to the early-time data before host-extinction correction. Prominent \ion{He}{1} features from unburnt He are predicted, which is consistent with the spectra of SN 2023zaw.  We note that the absence of broad \ion{O}{1} $\lambda$7774 and the presence of \ion{Ca}{2} NIR lines at early phases resemble the spectra of SN 2010X \citep{Kasliwal2010} which is a `.Ia' SN candidate (see Figure \ref{fig:spec_compare}). However, the fact that the decline of the $r$-band lightcurve slows down after $\sim$ 10 days post-peak, is not consistent with the `.Ia' SN models \citep{Shen2010}. %We cannot entirely rule out this scenario if the late-time lightcurve is powered by CSM interaction, evidence of which is seen in the late-time spectra.

\subsection{Accretion-induced collapse (AIC) of white dwarf}
O-Ne white dwarfs collapse to form a neutron star when it approaches Chandrasekhar mass. Numerical simulations predict very low explosion energy ($\sim 0.5 \times 10^{49}$ erg). We compare the bolometric lightcurve prediction from \citet{Darbha2010} with SN 2023zaw in Figure \ref{fig:modelcompare}. The predicted peak luminosity is $\sim$ 10 times fainter than that observed for SN 2023zaw. The models do not predict the late-time tail seen in SN 2023zaw. The nickel mass predicted in AIC models ($\lesssim 0.001$ \msun) is slightly smaller than that estimated for SN 2023zaw. The ejecta velocities predicted for AIC explosions are very high ($\sim0.1$c), which is not consistent with SN 2023zaw. 

%However, the nickel mass estimated for SN 2023zaw will be consistent if CSM interaction contributes to the bolometric luminosity.  
\subsection{White dwarf $-$ neutron star/black hole merger}
Spectroscopic models of WD-NS/BH mergers \citep{Gillanders2020} predict that the ejecta should be rich in intermediate elements such as O, Mg, S. This is not consistent with the observed spectra for SN 2023zaw. %Also, the nickel synthesised in these mergers is not sufficient to power the lightcurve

\subsection{Nickel free core-collapse supernova}
A potential powering mechanism for a fast-evolving, low-Ni mass SN is the `Ni-free' SN scenario \citep{Kleiser2014, Kleiser2018}. In this case, the supernova is not powered by decay of radionuclides like $^{56}$Ni but is instead powered by shock-interactions with an extended hydrogen-poor circumstellar medium (CSM). We compare the $r$-band lightcurve and bolometric luminosity of SN 2023zaw with Ni-free SN models from \citet{Kleiser2018} in Figure \ref{fig:modelcompare}. If there is a significant amount of Ni present ($> 0.05$ \Msun), it shows up as a second peak in the model lightcurves, which is not seen for SN 2023zaw. The models with 0 or 0.01 \Msun\ of nickel produce bright and short-lived peaks. However, none of the lightcurves can explain the rapid rise and the slower-evolving late-time lightcurve tail of SN 2023zaw. %We cannot rule out this scenario if the late-time lightcurve is powered by CSM interaction.

%\subsubsection{Core-collapse supernova from a low mass progenitor}

%The spectral template match to a Type Ib SN and the star-forming environment favor a massive star origin.

%\textit{Low mass end of core-collapse SN}

\subsection{Electron-capture SN}
\begin{figure*}
    \centering
    \includegraphics[width=8.5cm]{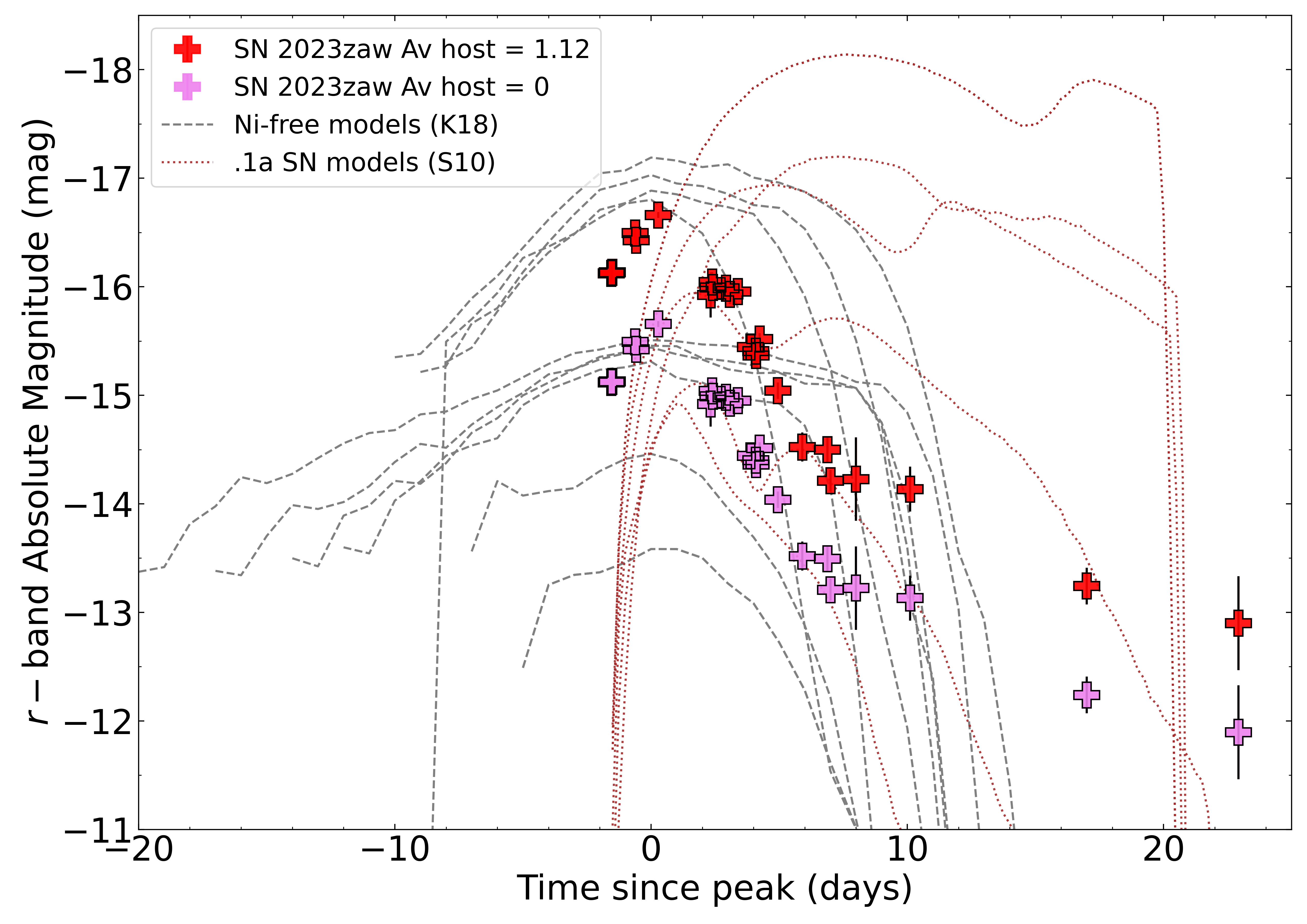}\includegraphics[width=8.5cm]{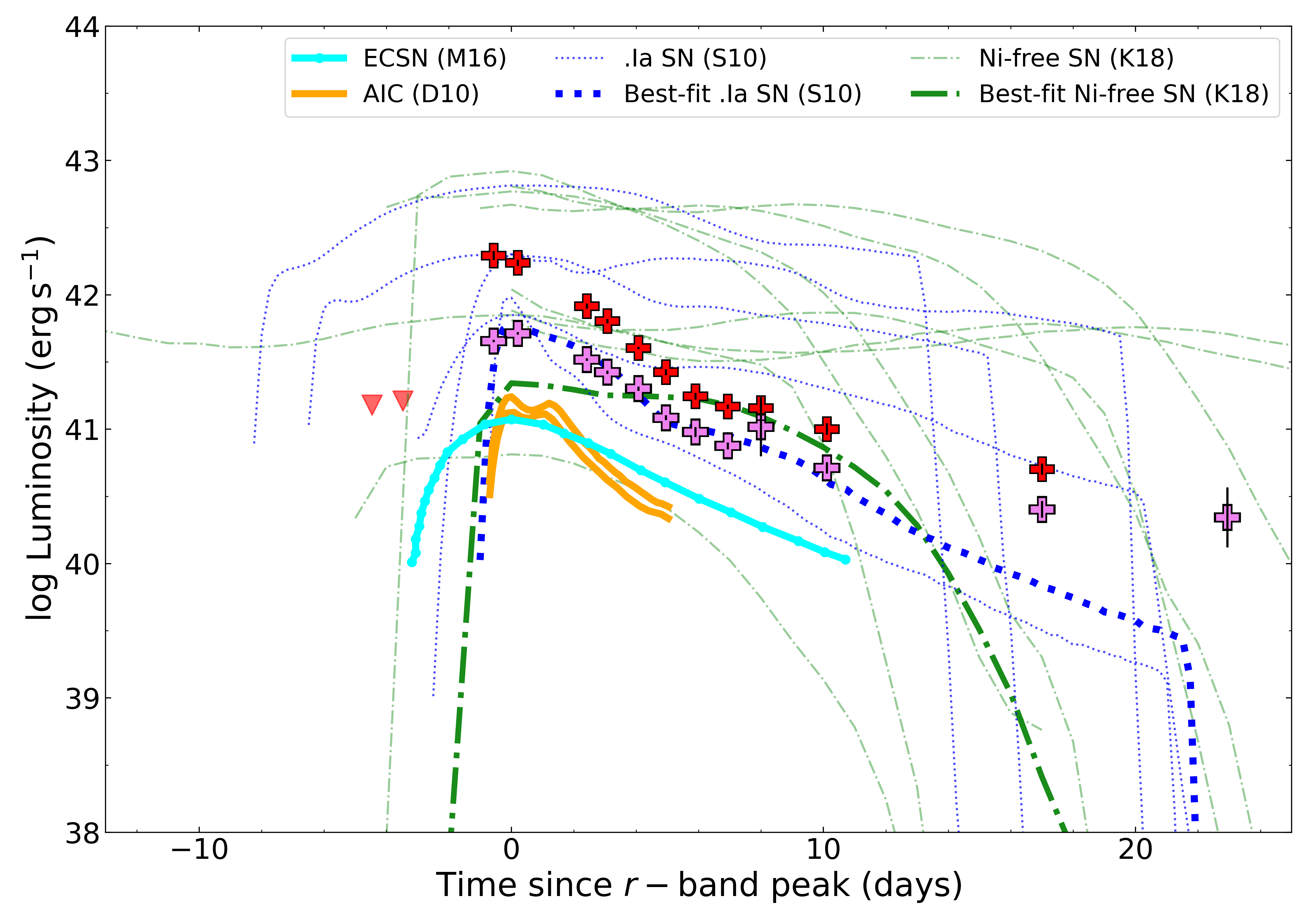}
    \caption{Comparison of the $r-$band and bolometric luminosity of SN 2023zaw with theoretical models for `.Ia' SNe \citep{Shen2010}, nickel-free SNe \citep{Kleiser2018}, electron-capture SN \citep{Moriya2016}, and accretion-induced collapse of white dwarf \citep{Darbha2010}. The bold blue and green lines indicate the best-fit `.Ia' and nickel-free SN models,  respectively.}
    \label{fig:modelcompare}
\end{figure*}

The low nickel mass is consistent with super Asymptotic Giant Branch (sAGB) progenitor stars with an O-Ne-Mg core that explodes as an electron-capture supernova (ECSN). However, a sAGB star has a hydrogen-rich envelope. Since, the spectra do not show hydrogen, a binary system is necessary, where the hydrogen has been stripped off through Roche-lobe overflow or common envelope ejection. Stripped-envelope ECSN models in \citet{Moriya2016} predict ejecta mass of $\sim0.3 - 0.6$ \msun, nickel mass of $\sim0.003$ \Msun, and explosion energy of $10^{50}$ erg, which is consistent with SN 2023zaw. Enhanced production of calcium is predicted for these SN, which is also consistent with the broad Ca emission lines seen in SN 2023zaw, as early as 2 days after peak. We compared the theoretical lightcurves from \citet{Moriya2016} with SN 2023zaw in Figure \ref{fig:modelcompare}. The predicted peak bolometric luminosities ($\sim 10^{41}\ \mathrm{erg\ s^{-1}}$) are $\sim10$ times fainter than the peak luminosity measured for SN 2023zaw. 
%It is possible that CSM interaction can boost the predicted luminosity.

\subsection{Ultra-stripped core-collapse supernova from a low-mass progenitor embedded in a He-rich CSM }

An ultra-stripped SN \citep{Tauris2015} is a possible channel for the formation of double neutron-star systems. They arise from a binary star system where a He-star 
undergoes a high degree of stripping by its smaller companion, such that the remaining ejecta mass is $<$ 1 \Msun. The nickel mass is also expected to be quite low ($\sim$ 0.01 \Msun). Due to the low nickel and ejecta mass, the lightcurves are expected to be faint and show fast evolution. The multiband lightcurve and bolometric properties of SN 2023zaw are consistent with an ultra-stripped SN scenario (Figures \ref{fig:phot_all}, \ref{fig:shockfit} and \ref{fig:rbandcompare_r}). We compare the measured ejecta and nickel mass for SN 2023zaw with other ultra-stripped SN candidates in the literature in Table \ref{table:ultracompare} and Figure \ref{fig:mnimej}. The estimated ejecta mass is comparable with other ultra-stripped SN candidates, while the nickel mass is a factor of $\sim5$ lower. We could also reproduce the observed lightcurve by using ultra-stripped SN progenitor models evolved with \mesa\ and exploded using \snec\ (see Section \ref{sec:radhydro}). 
%The measured envelope properties are consistent with the bound and unbound CSM properties of low-mass He stars (see Figure \ref{fig:bound}). From our radiation-hydrodynamic modeling with \snec, we find that the bound envelope extending to $\approx$ 50 \Rsun\ for a He-star with initial mass 2.6 \Msun\ provides a good fit to the observed lightcurve. Other models fail to reproduce either the early cooling-phase of the bolometric lightcurve (governed by the radius and mass in the bound envelope) or the late-time decline rate (governed by the total ejecta mass). We note that the models including the unbound CSM predicted by \citet{Wu2022b} are largely inconsistent with our data. Luminosities and durations can be similar to what we are observing if the explosion energy is adjusted to be low ($<0.1\times10^{51}$~erg). However, they show far too low expansion velocities during the first week after explosion compared to what we derive from our blackbody fitting and spectroscopic absorption line measurements. 
\subsubsection{Implication of narrow \ion{He}{1} lines in the late-time spectra}
The presence of narrow \ion{He}{1} emission lines in the late-time spectra at 30 and 60 days post-peak implies the presence of a He-rich dense CSM shell, which is consistent with the predictions of ultra-stripped SN models. %If we assume a shock velocity of $\sim$ 0.1c, the CSM extends to $\sim 7 \times 10^{15}$ cm. 
We note that we do not see prominent narrow \ion{He}{1} emission lines in the spectra taken prior to 9 days after peak. \textcolor{black}{It is likely the broad \ion{He}{1} lines from the ejecta dominate over the narrow CSM emission lines in the initial days. But if there is actually no CSM interaction till +8 days, we can constrain the CSM shell to be at $\sim 10^{15}$ cm, which is consistent with the short-period binary models from \citet{Wu2022b}.} While the mass transfer model in \citet{Wu2022b} assumes that the stellar material lost through the L2 point is distributed spherically, a torus-like distribution of the CSM is also likely \citep{Pejcha2016, Lu2023}. The viewing angle will strongly influence the observed spectral line widths. In this scenario, the early spectra do not show narrow lines as the CSM is engulfed in the SN ejecta, similar to the model described in \citet{Andrews2018} for iPTF14hls. Once the photosphere recedes, the CSM is accelerated by the collision of the ejecta to intermediate velocities of $\approx$ 1000 km $\mathrm{s^{-1}}$. \textcolor{black}{In this scenario, the absence of narrow lines in the ALFOSC spectrum taken at +0 days can be explained if the CSM is confined within $~10^{14}$ cm, which is consistent with the majority of binary mass transfer models in \cite{Wu2022b}.} We note that the CSM interaction does not show up as an additional bump in the optical lightcurve. This could be explained by a very low mass He-rich CSM, which is consistent with the estimated low progenitor and ejecta masses. If CSM interaction does contribute significantly to the luminosity, the nickel mass we estimate should be considered as an upper limit.

\subsubsection{Implication of very low nickel mass}

Observations and parametrized 1D models of massive star explosions suggest that typically $\sim0.05$ \Msun\ Ni should be ejected from a core-collapse SN. \citet{Anderson2019} and \citet{Meza2020} measured the Ni-mass for a sample of Type II and Type Ibc SNe in the literature and found that the Ni-mass synthesized in SESNe was significantly higher than for Type II SNe. The lowest Ni-mass in their Type Ibc sample was 0.015 \Msun, and 25\% of all Type II SNe has a Ni-mass below this limit. However, this might be due to an observational bias as low Ni-mass SESNe are harder to detect as they are likely to be faint as well as fast-evolving. The extremely low amount of Ni-mass in SN 2023zaw ($\sim$ 0.002 \Msun) could imply a fallback event, where most of the Ni produced in the core-collapse immediately fall into the newly formed black hole \citep{Turatto1998, Sollerman2002, Balberg2000}. However, the observed high velocities do not favor this scenario. 

The other explanation is that the progenitor's initial mass lies in the low-mass end of core-collapse SN progenitors, as stars with low-mass iron cores produce a low amount of Ni. \citet{Stockinger2020} and \citet{Sandoval2021} obtained $\sim0.002 $-$ 0.005$ \Msun\ Ni mass for a 9.6 \Msun\ progenitor. More recent 3D simulations \citep{Burrows2024} estimate 0.002 $-$ 0.006 \Msun\ Ni-mass for a 9 \Msun\ ZAMS mass progenitor and $>$ 0.01 \Msun\ for progenitors with initial mass greater than 9.25 \Msun. We note that these models are not for stripped-envelope SNe. However, if they undergo mass-loss after He-core burning, the core evolution of a star is decoupled from the envelope evolution and the nickel mass is expected to be the same. %\citet{Ertl2020} exploded He-stars in binary systems and estimated 0.02 \Msun\ Ni-mass for a 2.7 \Msun\ He-star initial mass. 
The amount of Ni synthesized in ultra-stripped models for low-mass CO stars is $\lesssim$ 0.01 \Msun\ \citep{Moriya2017, Sawada2022}, which is consistent with our estimate for SN 2023zaw. %While some of these models are for single stars, the nickel mass estimates are likely also hold for binary evolution. Generally, the mass-transfer in binary systems occur after He-core burning and the core and envelope evolution are de-coupled.

\section{Conclusion}
\label{sec:summary}

In this paper, we have presented the observations and analysis of an unusual stellar explosion $-$ SN 2023zaw. It has the lowest nickel mass among all stripped-envelope SNe discovered so far. SN 2023zaw also shows the fastest evolution among all stripped-envelope SNe, with a time above half maximum of 4.9 days in the $r-$band. The photospheric spectra show broad \ion{He}{1} emission lines. It also shows broad Ca NIR emission lines as early as 2 days after peak. 

We performed radiation-hydrodynamical and analytical modeling of the lightcurve by fitting with a combination of shock-cooling emission and nickel decay. The early-time lightcurve fits well with shock-cooling emission, with an envelope mass of $\approx 0.2$\ \Msun and an envelope radius of $\approx 50$ \Rsun. The estimated ejecta and nickel masses are $\approx 0.5$ \Msun\ and $\approx 0.002$ \Msun\ respectively. 

We interpret SN 2023zaw as an ultra-stripped SN from a low-mass He-star $-$ compact object binary. The late-time evolution of the He-star results in an extended He-rich envelope. There is also spectral evidence for thick He-rich CSM, likely from pre-SN mass loss episodes from late-time Roche lobe overflow. The very low nickel mass, ejecta mass, and explosion energy can only be explained by a progenitor with an initial mass less than around 10 \Msun.

The late-time spectra show prominent narrow \ion{He}{1} at a FWHM velocity of $\sim1000$ $\mathrm{km\ s^{-1}}$, indicating the presence of He-rich CSM. It is unclear if CSM interaction also provides a powering mechanism for SN 2023zaw. A number of other models for faint and rapidly evolving transients such as the nickel-free SN, accretion induced collapse, electron capture supernova were ruled out based on the peak luminosity and the presence of a late-time tail in the lightcurve of SN 2023zaw. The excess luminosity might be explained if the luminosity is boosted by CSM interaction. In this case, the estimated nickel mass should be treated as an upper limit. Including CSM interaction modeling in the future will help conclusively favor or rule out these scenarios. Also, late-time multi-wavelength follow-up will provide useful diagnostics in determining the CSM properties, the pre-SN mass-loss mechanism, and the evolutionary channel \citep{Matsuoka2020, Kashiyama2022}. 

Spectroscopic modeling is currently missing for most of the theoretical scenarios for producing faint and rapid transients. Detailed modeling is required to explain some unusual  spectroscopic features of SN 2023zaw such as $-$ the presence of strong Ca NIR lines close to peak, redshifted component of \ion{He}{1} $\lambda5876$ emission line, not reaching nebular phase till 30 days in spite of the low ejecta mass, the emergence of narrow \ion{He}{1} emission lines and the absence of [\ion{Ca}{2}], [\ion{O}{1}] and Ca NIR emission lines in the nebular phase.  %, which will be critical in inferring the evolutionary channel. %.Detailed spectroscopic modeling is required to explain these features in a consistent picture. 

The extremely low nickel mass of SN 2023zaw gives rise to the question $-$ what is the lowest nickel mass that is synthesized in a stripped-envelope SN? 
Current literature shows that the nickel mass measured in stripped-envelope SNe is significantly higher than in Type II SNe \citep{Anderson2019}. It is likely that this is due to an observational bias, as SESNe with low nickel masses are likely to be faint and fast evolving and hence harder to detect, classify, and follow-up. SN 2023zaw underscores the existence of this undiscovered population of low nickel mass SESNe ($< 0.005$ \msun) with progenitor masses in the low mass end of core-collapse SNe ($< 10$ \msun).  A systematic study of the lowest-nickel mass stripped-envelope SNe, including their rates and the nickel mass distribution of low-luminosity SESNe will be explored in future work. Candidates for such low nickel mass stripped-envelope SNe include double peaked Type Ibc SNe such as SN 2021inl \citep{Jacobson2022b, Das2023b}, rapidly-evolving ($\mathrm{t_{1/2,r}} < 10$ days) Type IIb SNe \citep[e.g.,][Fremling et al. in prep]{Ho2023, Das2023a}, rapidly-declining H-poor SNe such as SN 2020ghq (Wang et al. in prep).  For SNe with extremely low Ni-mass, the shock cooling emission is likely to dominate over nickel decay. The ultra-stripped SNe candidates in the literature show an early luminosity excess that is powered by shock-cooling. Future wide-field UV surveys such as ULTRASAT \citep{Sagiv2014, Shvartzvald2023}, UVEX \citep{Kulkarni2021}, and deep ground-based surveys such as LSST \citep{Ivezic08} in synergy with high-cadenced surveys such as ZTF will provide an exciting opportunity to explore this phase space.

%Also, there exists rapidly-evolving ($\mathrm{t^{1/2}_r} < 10$ days) Type IIb SNe, where the primary peak is likely powered by shock-cooling emission, due to very low nickel mass \citep[e.g.,][Fremling et al. in prep]{Ho2023, Das2023a}.

%Other stripped-envelope SN with very low nickel mass includes the double-peaked Type Ib SN $-$ 2021inl \citep{Jacobson2022b, Das2023b}, Type IIb SNe like $-$ SN 2021sjt, SN 2019pof \citep{Das2023a}. These SNe are powered by shock-cooling emission during the early phases followed by radioactivity. 
%Current systematic, deep and high-cadenced surveys such as the CLU and BTS experiments and future surveys such as LSST provide an exciting opportunity to explore this phase space. The rates of such events and the distribution of nickel mass in the low mass end of SESNe will be explored in future work. 

\section{Data availability}
All the photometric and spectroscopic data used in this work will be made available on \href{https://github.com/kaustavkdas/SN2023zaw}{GitHub} after publication.

The optical photometry and spectroscopy will also be made public through WISeREP, the Weizmann Interactive Supernova Data Repository \citep{Yaron2012}.

\section{Acknowldegement}
KKD acknowledges support from the Schwartz Reisman Collaborative Science Program, which is supported by the Gerald Schwartz and Heather Reisman Foundation.

M.W.C acknowledges support from the National Science Foundation with grant numbers PHY-2308862 and PHY-2117997.

S. Schulze is partially supported by LBNL Subcontract NO.\ 7707915. 

Based on observations obtained with the Samuel Oschin Telescope 48-inch and the 60-inch Telescope at the Palomar Observatory as part of the Zwicky Transient Facility project. ZTF is supported by the National Science Foundation under Grant No. AST-2034437 and a collaboration including Caltech, IPAC, the Oskar Klein Center at Stockholm University, the University of Maryland, University of California, Berkeley, the University of Wisconsin at Milwaukee, University of Warwick, Ruhr University, Cornell University, Northwestern University and Drexel University. Operations are conducted by COO, IPAC, and UW.

SED Machine is based upon work supported by the National Science Foundation under
Grant No. 1106171. 

The ZTF forced-photometry service was funded under the Heising-Simons Foundation grant \#12540303 (PI: Graham).

The Gordon and Betty Moore Foundation, through both the Data-Driven Investigator Program and a dedicated grant, provided critical funding for SkyPortal.

This research has made use of the NASA/IPAC Extragalactic Database (NED), which is funded by the National Aeronautics and Space Administration and operated by the California Institute of Technology.

The data presented here were obtained in part with ALFOSC, which is provided by the Instituto de Astrofisica de Andalucia (IAA) under a joint agreement with the University of Copenhagen and NOT.

The Liverpool Telescope is operated on the island of La Palma by Liverpool John Moores University in the Spanish Observatorio del Roque de los Muchachos of the Instituto de Astrofisica de Canarias with financial support from the UK Science and Technology Facilities Council.
Based on observations made with the Italian Telescopio Nazionale Galileo (TNG) operated on the island of La Palma by the Fundación Galileo Galilei of the INAF (Istituto Nazionale di Astrofisica) at the Spanish Observatorio del Roque de los Muchachos of the Instituto de Astrofisica de Canarias.

The W. M. Keck Observatory is operated as a scientific partnership among the California Institute of Technology, the University of California and the National Aeronautics and Space Administration. The Observatory was made possible by the generous financial support of the W. M. Keck Foundation. The authors wish to recognize and acknowledge the very significant cultural role and reverence that the summit of Maunakea has always had within the indigenous Hawaiian community.  We are most fortunate to have the opportunity to conduct observations from this mountain.

MMT Observatory access was supported by Northwestern University and the Center for Interdisciplinary Exploration and Research in Astrophysics (CIERA).

\FloatBarrier
\bibliography{main}
\bibliographystyle{aasjournal}

\appendix
\newcounter{Atable}
\setcounter{Atable}{0}
\newcounter{Afigure}
\setcounter{Afigure}{0}

\renewcommand\thefigure{\Alph{section}\arabic{Afigure}} 
\renewcommand\thetable{\Alph{section}\arabic{Atable}}

\section{Timescale - Luminosity phase space}
\addtocounter{Afigure}{1}
\addtocounter{Atable}{1}

\begin{figure}[H]
    \centering
    \includegraphics[width=8.5cm]{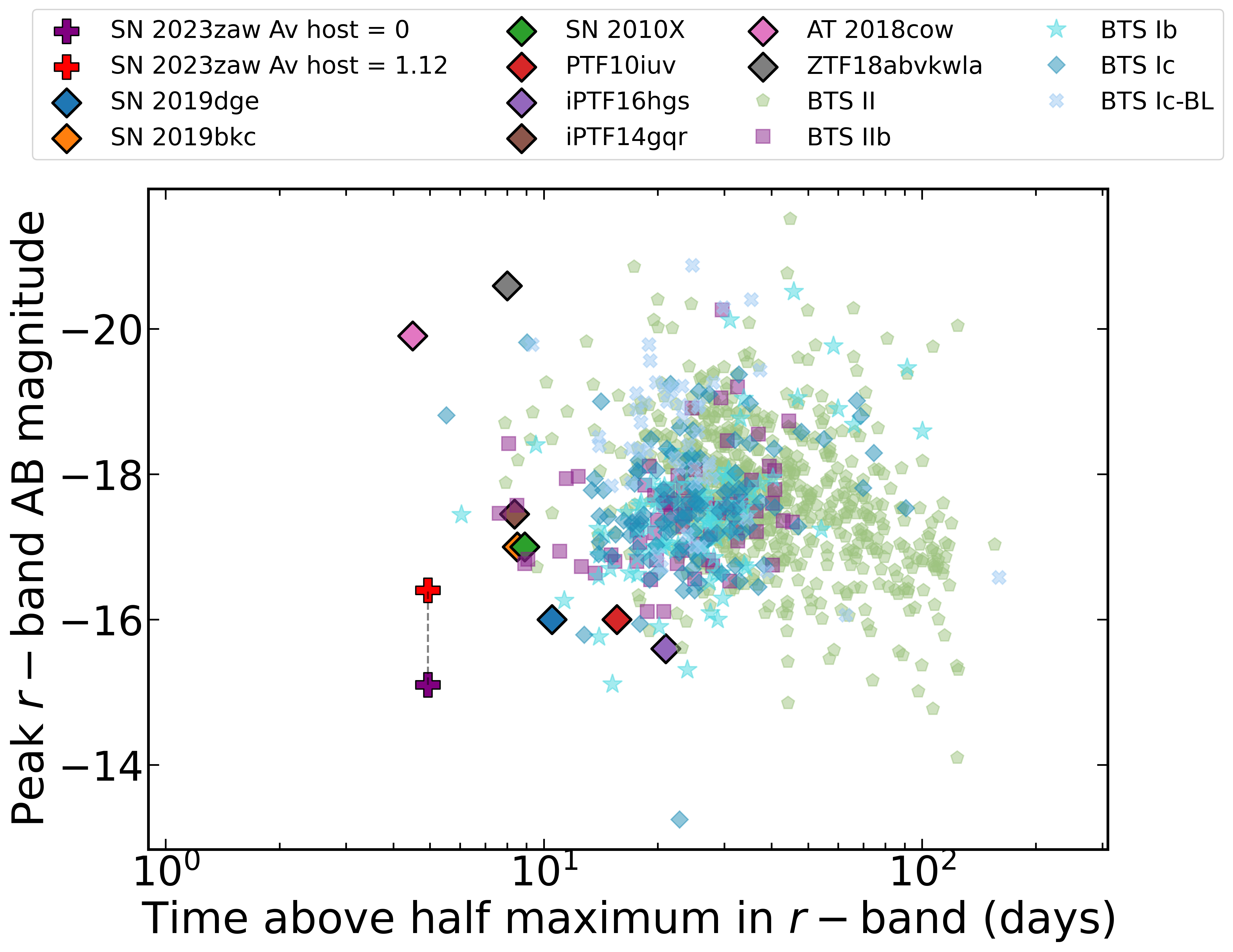}
    \caption{We compare the time above half maximum and peak luminosity of SN 2023zaw with stripped-envelope supernovae from the BTS experiment \citep{Perley2020}. For sources with two peaks in the $r-$band lightcurve, the second peak is used. We also compare with other fast-evolving transients $-$ SN 2019dge \citep{Yao2020}, SN 2019bkc \citep{Prentice2020, Chen2020}, iPTF14gqr \citep{De2018c}, iPTF16hgs \citep{De2018d}, AT 2018cow \citep{Perley2019b}, ZTF18abvkwla or AT2023abbd \citep{Ho2020}.}
    \label{fig:timescale}
\end{figure}

In Figure \ref{fig:timescale}, we compare the time above half maximum and peak luminosity of SN 2023zaw with stripped-envelope supernovae and other fast-evolving transients in the literature with rise and decline time constraints.

\section{Host galaxy emission lines}
Table \ref{tab:explosionsite} shows the emission line flux measurements of the star-forming region at the SN site.
%\begin{figure}[H]
%    \centering
%    \includegraphics[width=6cm]{23zaw_galaxy.png}\includegraphics[width=10cm]{SN2023zaw_KCWI_Halpha.pdf}
%    \caption{\textit{Left:} Pan-STARRS image of UGC 03048, the host-galaxy of SN 2023zaw. The location of the transient is indicated by the white cross at the center. \textit{Right:} Star-formation rate map of the environment around SN\,2023zaw. SN\,2023zaw (position marked by the red circle) exploded in a star-forming region of the host galaxy close to regions of more vigorously star-forming regions. The image has a size of $4.1\times6.7$~kpc. The SFR scale is not corrected for attenuation.}
%    \label{fig:galaxy}
%\end{figure}

\begin{table}[H]
\caption{Emission line flux measurements of the star-forming region at the SN site.}\label{tab:explosionsite}
\centering
\begin{tabular}{cc}
\hline\hline
Transition                 & Flux \\
                           & $\left(10^{-17}\,{\rm erg\,cm}^{-2}\,{\rm s}^{-1}\right)$\\
\hline
{[\ion{O}{2}]}\,$\lambda\lambda$\,3726,3729	&$	61.10	\pm	5.23	$\\
{[\ion{O}{3}]}\,$\lambda$\,4959	&$	4.63	\pm	0.15	$\\
{[\ion{O}{3}]}\,$\lambda$\,5007	&$	13.90	\pm	0.46	$\\
H$\gamma$                	        &$	14.83	\pm	1.20	$\\
H$\beta$                 	        &$	42.48	\pm	2.51	$\\
{[\ion{N}{2}]}\,$\lambda$\,6549	&$	26.04	\pm	1.27	$\\
H$\alpha$	                        &$	203.98	\pm	6.74	$\\
{[\ion{N}{2}]}\,$\lambda$\,6584	&$	78.12	\pm	3.81	$\\
{[\ion{S}{2}]}\,$\lambda$\,6718	&$	40.84	\pm	1.61	$\\
{[\ion{S}{2}]}\,$\lambda$\,6732	&$	28.69	\pm	1.43	$\\
\hline\hline
\end{tabular}
\tablecomments{
No measurement is corrected for reddening.
}
\end{table}

\newpage

\section{Photometry Data}
A truncated version of the photometry table is shown in Table \ref{tab:all_phot}. All the photometric and spectroscopic data used in this work will be made available on \href{https://github.com/kaustavkdas/SN2023zaw}{GitHub} after publication.

\vspace{1mm}

%The optical photometry and spectroscopy will also be made public through WISeREP, the Weizmann Interactive Supernova Data Repository \citep{Yaron2012}.
\begin{table}[H] 
\begin{center} 
\caption{Truncated photometry data for SN 2023zaw. The full machine-readable version will be made available on \href{https://github.com/kaustavkdas/SN2023zaw}{GitHub} after publication. The photometry data has been corrected for Milky Way extinction.}

\begin{tabular}{ccccc} 
\hline 

MJD &  Phase$^{\textcolor{blue}{a}}$  & Instrument & Filter &  AB Magnitude \\ \\ \hline 

$60285$ & $-2$ & ATLAS & $o$ & $19.33 \pm 0.26$  \\  

$60285$ & $-2$ & P48 & $g$ & $18.34 \pm 0.06$  \\  

$60285$ & $-2$ & ATLAS & $o$ & $18.16 \pm 0.07$  \\  

$60285$ & $-2$ & SEDM & $r$ & $18.09 \pm 0.07$  \\  

$60285$ & $-2$ & SEDM & $g$ & $18.23 \pm 0.05$  \\  

$60285$ & $-2$ & SEDM & $r$ & $18.08 \pm 0.04$  \\  

$60285$ & $-2$ & SEDM & $i$ & $17.91 \pm 0.06$  \\  

$60286$ & $-1$ & SEDM & $r$ & $17.72 \pm 0.06$  \\  

$60286$ & $-1$ & SEDM & $g$ & $17.98 \pm 0.05$  \\  

$60286$ & $-1$ & SEDM & $r$ & $17.78 \pm 0.04$  \\  

$60286$ & $-1$ & SEDM & $i$ & $17.65 \pm 0.02$  \\  

$60286$ & $-1$ & ATLAS & $c$ & $17.88 \pm 0.05$  \\  

$60287$ & $0$ & P48 & $g$ & $17.96 \pm 0.05$  \\  

$60287$ & $0$ & P48 & $g$ & $17.97 \pm 0.04$  \\ \hline

\end{tabular}  \label{tab:all_phot} 
\end{center} 
\footnotesize $^a$ Days since $r$-band peak (MJD = 60287.7).
\end{table}

\vspace{1mm}

\section{Spectroscopy Collage}

\begin{figure}[H]
    \centering
    \includegraphics[width=15.5cm]{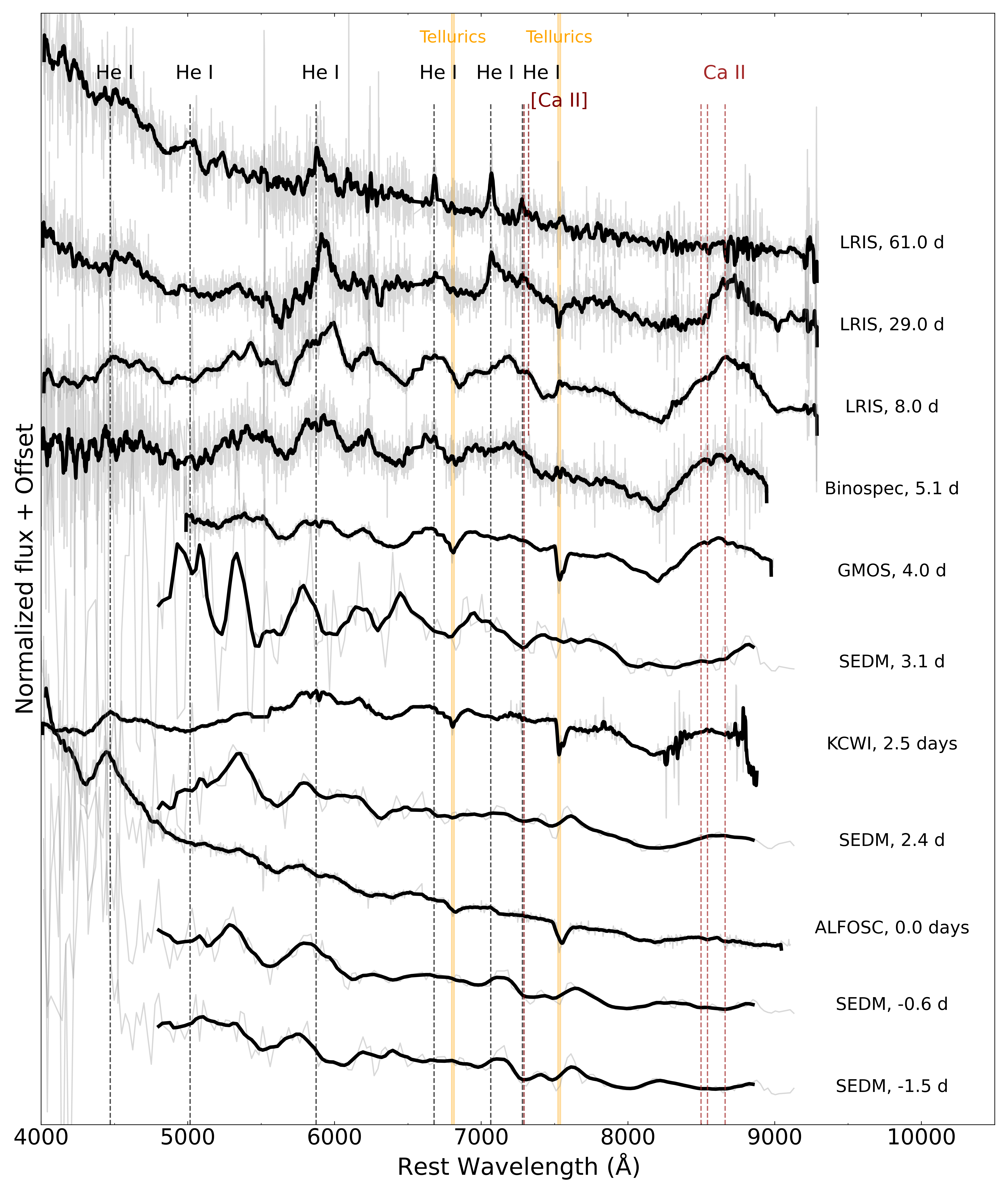}
    \caption{Spectral sequence for SN 2023zaw. The phases are in rest-frame days since the $r$-band peak (MJD = 60287.7). See Section \ref{section:spectra} for details on the obtained spectra.}
    \label{fig:spec_all}
\end{figure}

A collage all the spectra of SN 2023zaw is shown in Figure \ref{fig:spec_all}.

\section{Spectroscopy Log}
The spectroscopy log is shown in Table \ref{tab:spec_log}.
\begin{table}[H] 
\begin{center} 
\caption{Spectroscopy log.} 
\begin{tabular}{ccc} 
\hline 

Date &  Phase$^{\textcolor{blue}{a}}$  & Instrument \\ \\ \hline 

Dec 12 2023 & $-1.5$ & SEDM   \\  

Dec 08 2023 & $-0.6$ & SEDM \\  

Dec 09 2023 & $0.0$ & ALFOSC   \\  

Dec 11 2023 & $+2.4$ & SEDM \\  

Dec 12 2023 & $+2.5$ & KCWI   \\  

Dec 12 2023 & $+3.1$ & SEDM   \\  

Dec 13 2023 & $+4.0$ & GMOS   \\  

Dec 14 2023 & $+5.1$ & Binospec   \\  

Dec 17 2023 & $+8.0$ & LRIS   \\  

Jan 07 2024 & $+29.0$ & LRIS  \\ 

Feb 07 2024 & $+61.0$ & LRIS  \\ \hline   

\end{tabular}  \label{tab:spec_log} 
\end{center} 
\footnotesize $^a$ Days since $r$-band peak (MJD = 60287.7).
\end{table}

\section{Color Evolution Plot}
\begin{figure}[H]
    \centering
    \includegraphics[width=8.8cm]{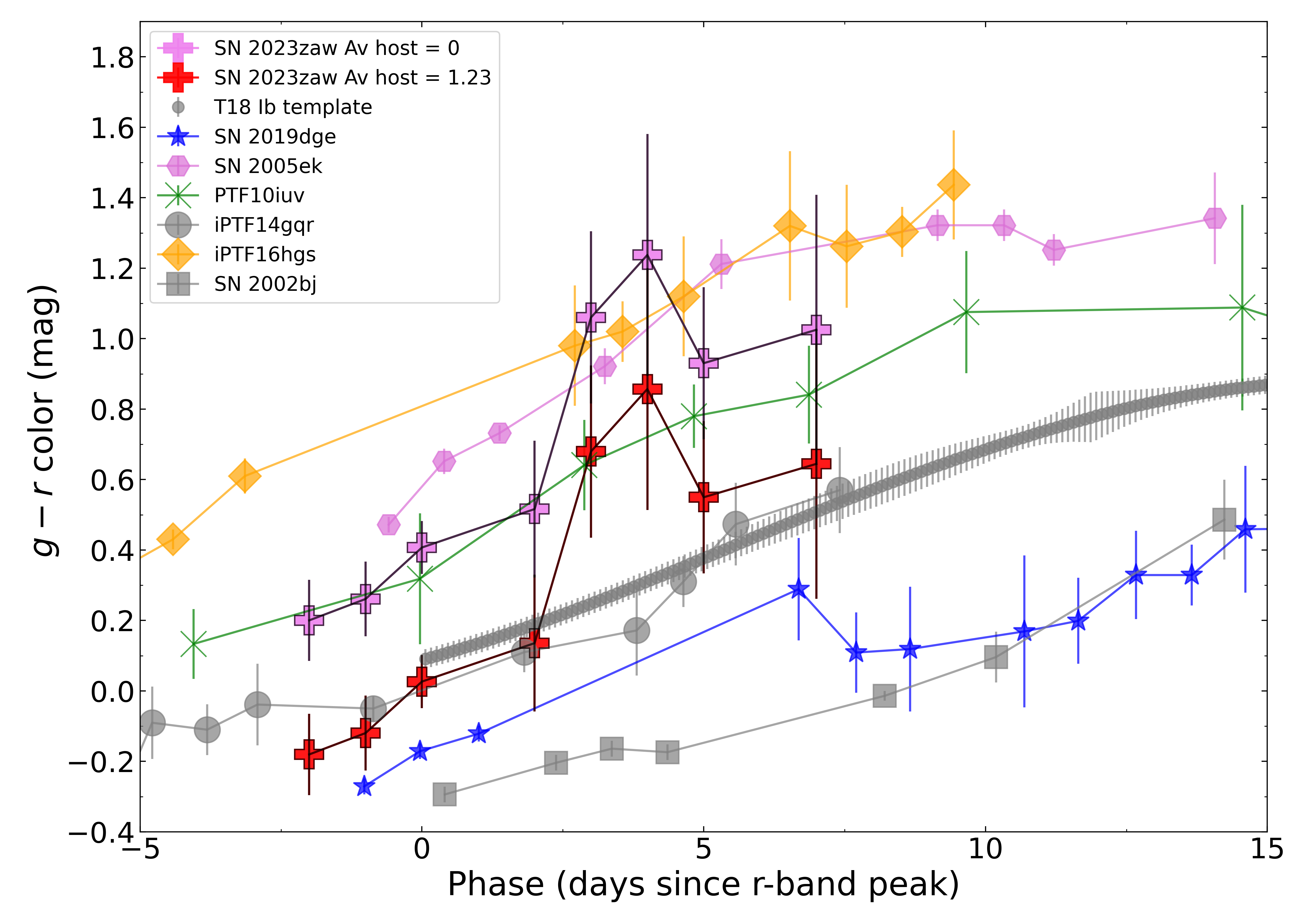}\includegraphics[width=9.8cm]{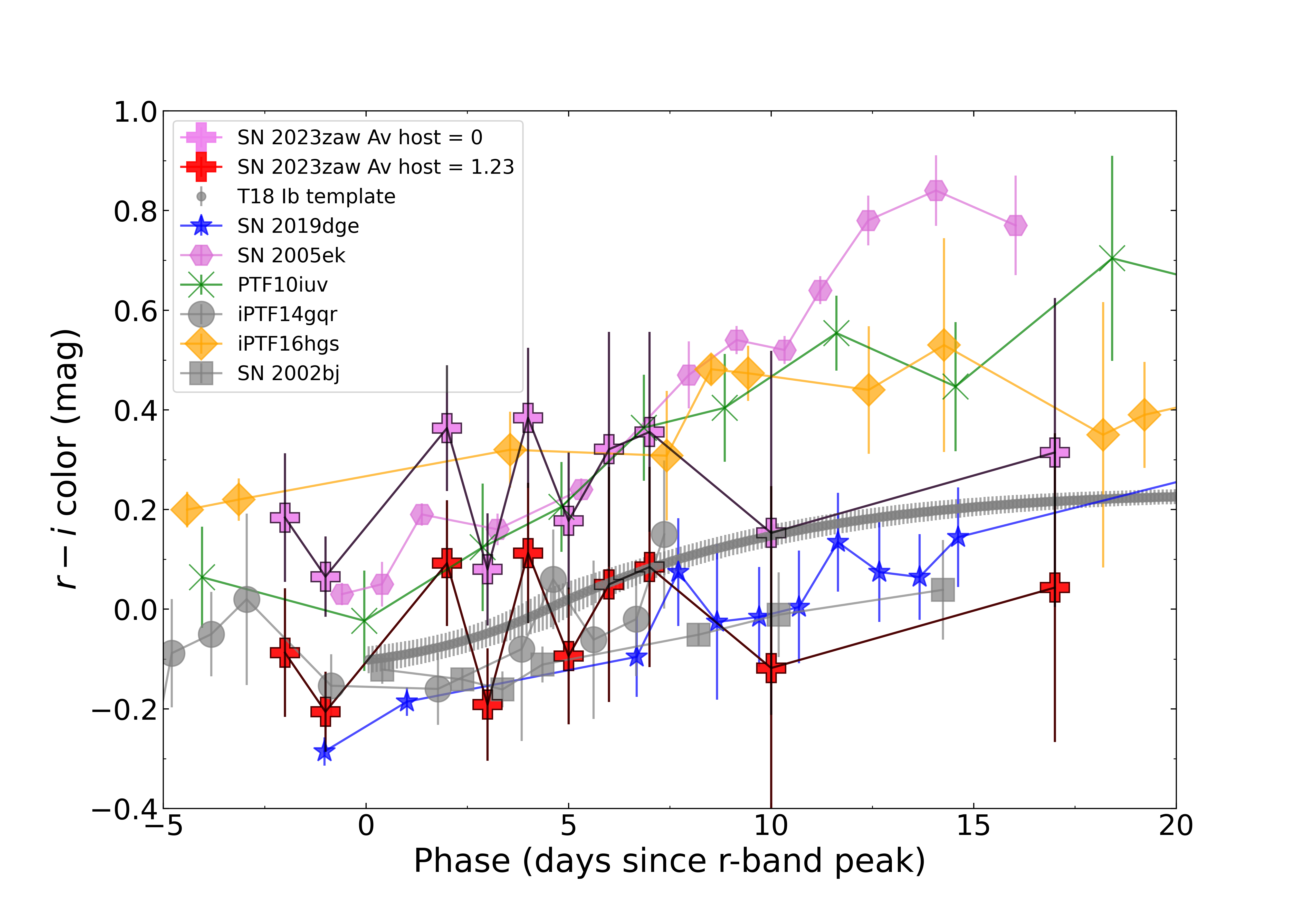}
    \caption{The $g-r$ (\textit{left}) and $r-i$ (\textit{right}) color evolution of SN 2023zaw compared to other fast transients $-$ SN 2005ek \citep{Drout2013}, SN 2019dge \citep{Yao2020}, iPTF14gqr \citep{De2018c}, iPTF14hgs \citep{De2018d}. The gray line shows the intrinsic color template for Type Ib SNe from \citet{Stritzingetr2018}.}
    \label{fig:color}
\end{figure}

The $g-r$ and $r-i$ color evolution of SN 2023zaw compared to other SNe in the literature is shown in Figure \ref{fig:color}.

\section{Lightcurve comparison}

\begin{figure}[H]
    \centering
    \includegraphics[width=9.5cm]{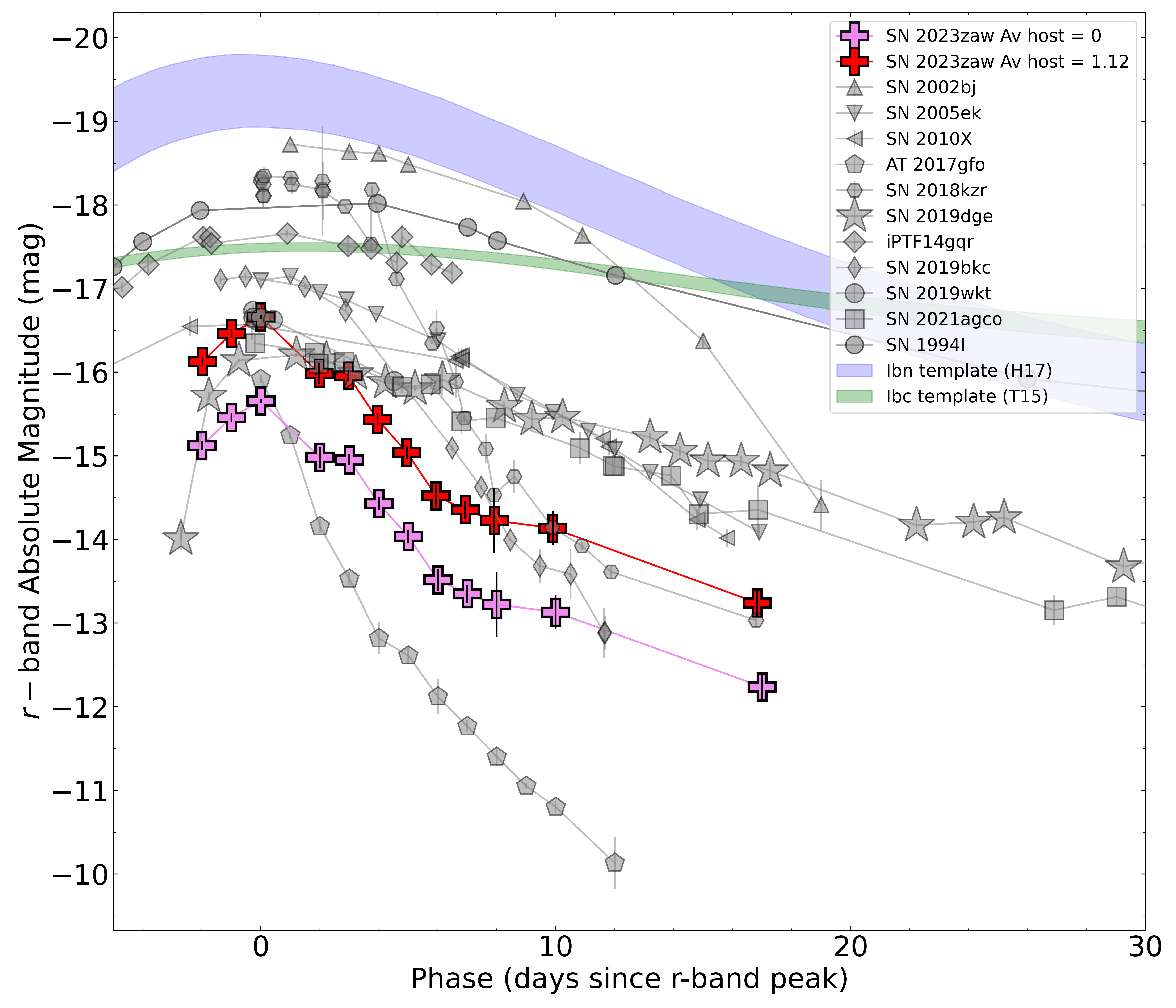}%\includegraphics[width=9.5cm]%{23zaw_g_collage.png}
    \caption{The $r-$band lightcurve of SN 2023zaw compared with other fast-evolving transients in the literature $-$ fast-declining Type I SNe $-$ SN 2010X \citep{Kasliwal2010}, SN 2019bkc \citep{Prentice2020, Chen2020}, the kilonova AT 2017gfo \citep{Abbott2017}, ultra-stripped SN candidates $-$ SN 2019dge \citep{Yao2020}, iPTF14gqr \citep{De2018c}, SN 2019wxt \citep{Shivkumar2023, Agudo2023}, SN 2022agco \citep{Yan2023}.}
 \label{fig:rbandcompare_r}
   
\end{figure}

In Figure \ref{fig:rbandcompare_r}, we compare the $r-$band lightcurve of SN 2023zaw with other fast-evolving transients in the literature.

\section{Blackbody best-fit parameters}
The summary of the best-fit blackbody properties can be found in Table \ref{table:bb_table}.

\begin{table}[H] 
\begin{center} 
\caption{Summary of the blackbody properties for SN 2023zaw. A machine-readable version of this table will be made available on \href{https://github.com/kaustavkdas/SN2023zaw}{GitHub} after publication.} 
\begin{tabular}{cccc} 
\hline 

Phase &  Log Luminosity   &  Temperature  & Radius  \\
(days) & ($\mathrm{erg\ s^{-1}}$) & (K) & ($\Rsun$) \\ \hline

$-2$ & $42.40^{+0.10}_{-0.08}$ & $15620^{+1850}_{-1440}$ & $3500^{+360}_{-360}$ \\ 

$-1$ & $42.29^{+0.03}_{-0.03}$ & $11050^{+510}_{-490}$ & $6180^{+340}_{-320}$ \\ 

$0$ & $42.24^{+0.02}_{-0.02}$ & $8960^{+500}_{-460}$ & $8820^{+760}_{-700}$ \\ 

$2$ & $41.92^{+0.02}_{-0.02}$ & $6900^{+340}_{-330}$ & $10270^{+860}_{-750}$ \\ 

$3$ & $41.80^{+0.02}_{-0.02}$ & $6710^{+300}_{-280}$ & $9530^{+710}_{-670}$ \\ 

$4$ & $41.60^{+0.01}_{-0.01}$ & $5520^{+230}_{-210}$ & $11190^{+880}_{-830}$ \\ 

$5$ & $41.42^{+0.02}_{-0.02}$ & $6100^{+380}_{-340}$ & $7440^{+790}_{-720}$ \\ 

$6$ & $41.24^{+0.02}_{-0.03}$ & $5060^{+470}_{-430}$ & $8780^{+1710}_{-1360}$ \\ 

$7$ & $41.17^{+0.02}_{-0.02}$ & $5370^{+320}_{-290}$ & $7140^{+870}_{-760}$ \\ 

$8$ & $41.16^{+0.07}_{-0.08}$ & $5160^{+10}_{-10}$ & $7640^{+590}_{-660}$ \\ 

$10$ & $41.00^{+0.04}_{-0.05}$ & $5070^{+10}_{-10}$ & $6630^{+310}_{-330}$ \\ 

$17$ & $40.70^{+0.04}_{-0.04}$ & $5000^{+10}_{-10}$ & $4830^{+220}_{-240}$ \\   \hline

\end{tabular}  \label{table:bb_table} 
\end{center} 
\end{table}

%\begin{table}[H] 
%\begin{center} 
%\caption{Summary of the blackbody properties for SN 2023zaw. A machine-readable version of this table will be made available on \href{https://github.com/kaustavkdas/SN2023zaw}{GitHub} after publication.} 
%\begin{tabular}{cccc} 
%\hline 

%Phase &  Log Luminosity   &  Temperature  & Radius  \\
%(days) & ($\mathrm{erg\ s^{-1}}$) & (K) & ($\Rsun$) \\ \hline

%$-2$ & $42.52^{+0.12}_{-0.10}$ & $17150^{+2260}_{-1730}$ & $3330^{+360}_{-350}$ \\ 

%$-1$ & $42.37^{+0.04}_{-0.04}$ & $11740^{+620}_{-540}$ & $5970^{+330}_{-330}$ \\ 

%$0$ & $42.29^{+0.03}_{-0.02}$ & $9330^{+580}_{-510}$ & $8670^{+770}_{-730}$ \\ 

%$2$ & $41.95^{+0.02}_{-0.02}$ & $7110^{+370}_{-350}$ & $10110^{+860}_{-760}$ \\ 

%$3$ & $41.84^{+0.02}_{-0.02}$ & $6920^{+320}_{-300}$ & $9370^{+710}_{-650}$ \\ 

%$4$ & $41.63^{+0.01}_{-0.01}$ & $5670^{+240}_{-220}$ & $10980^{+890}_{-810}$ \\ 

%$5$ & $41.46^{+0.03}_{-0.02}$ & $6290^{+410}_{-370}$ & $7270^{+800}_{-700}$ \\ 

%$6$ & $41.27^{+0.03}_{-0.03}$ & $5190^{+520}_{-460}$ & $8600^{+1730}_{-1390}$ \\ 

%$7$ & $41.20^{+0.02}_{-0.02}$ & $5510^{+340}_{-320}$ & $7030^{+870}_{-770}$ \\ 

%$8$ & $41.26^{+0.26}_{-0.11}$ & $4480^{+4060}_{-1330}$ & $10400^{+14170}_{-6900}$ \\ 

%$10$ & $41.13^{+0.12}_{-0.08}$ & $6970^{+1710}_{-1260}$ & $4040^{+1570}_{-1090}$ \\ 

%$17$ & $40.84^{+0.45}_{-0.10}$ & $7560^{+6750}_{-2260}$ & $2460^{+2000}_{-1300}$ \\ \hline

%$23$ & $40.80^{+2.09}_{-0.23}$ & $8240^{+57530}_{-4200}$ & $1840^{+4560}_{-1490}$ \\ 

%$27$ & $40.83^{+0.95}_{-0.27}$ & $3340^{+6120}_{-1620}$ & $9780^{+86040}_{-8240}$ \\ \hline

%\end{tabular}  \label{table:bb_table} 
%\end{center} 
%\end{table}

\section{Comparison of envelope properties}
\begin{figure}[H]
    \centering
    \includegraphics[width=8.5cm]{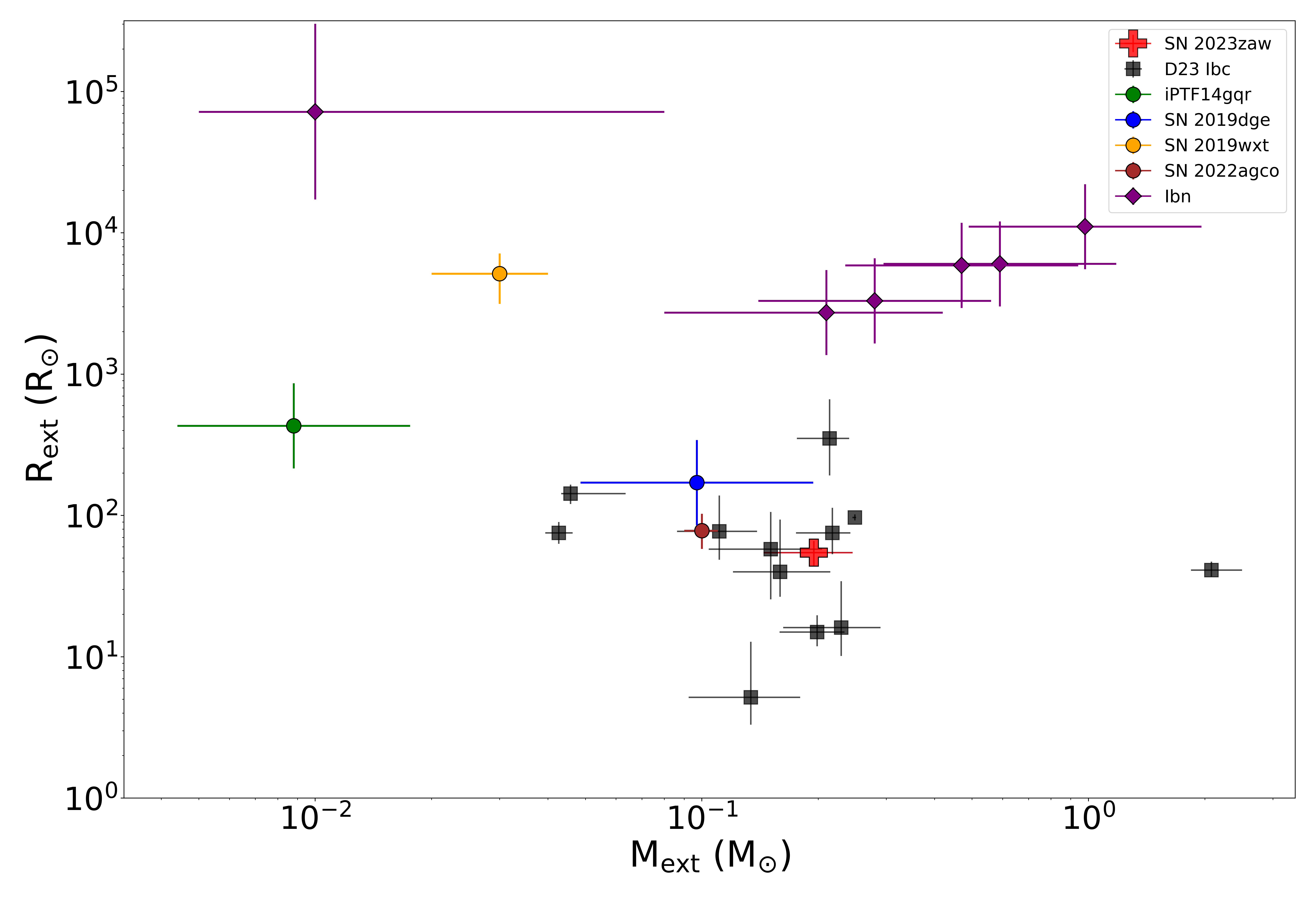}
    \caption{The envelope properties of SN 2023zaw (red cross) compared with double-peaked Type Ibc SNe (black squares) from \citet{Das2023b}, USSN candidates (circles) $-$ iPTF14gqr, SN 2019dge, SN 2019wxt, SN 2022agco \citep{De2018c, Yao2020, Shivkumar2023, Agudo2023, Yan2023}, and Type Ibn SNe (purple diamonds) $-$ SN 2006jc, SN 2019wep, SN 2019up, SN 2012jpk, SN 2019deh, LSQ13ddu \citep{Anupama2009, Pellegrino2022b, Clark2020}.}
    \label{fig:envelope}
\end{figure}

\begin{figure}[H]
    \centering
    \includegraphics[width=8.5cm]{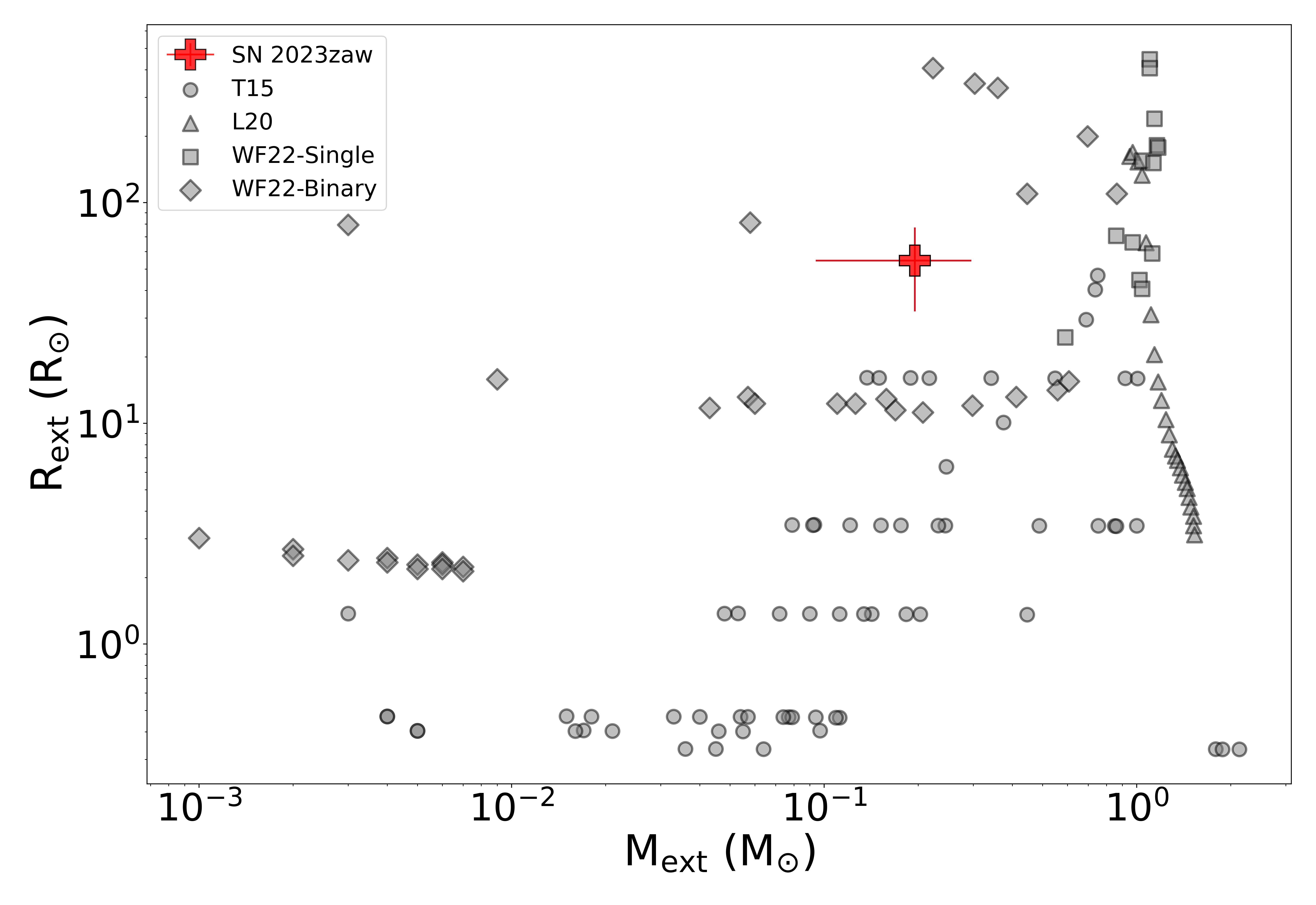}\includegraphics[width=8.5cm]{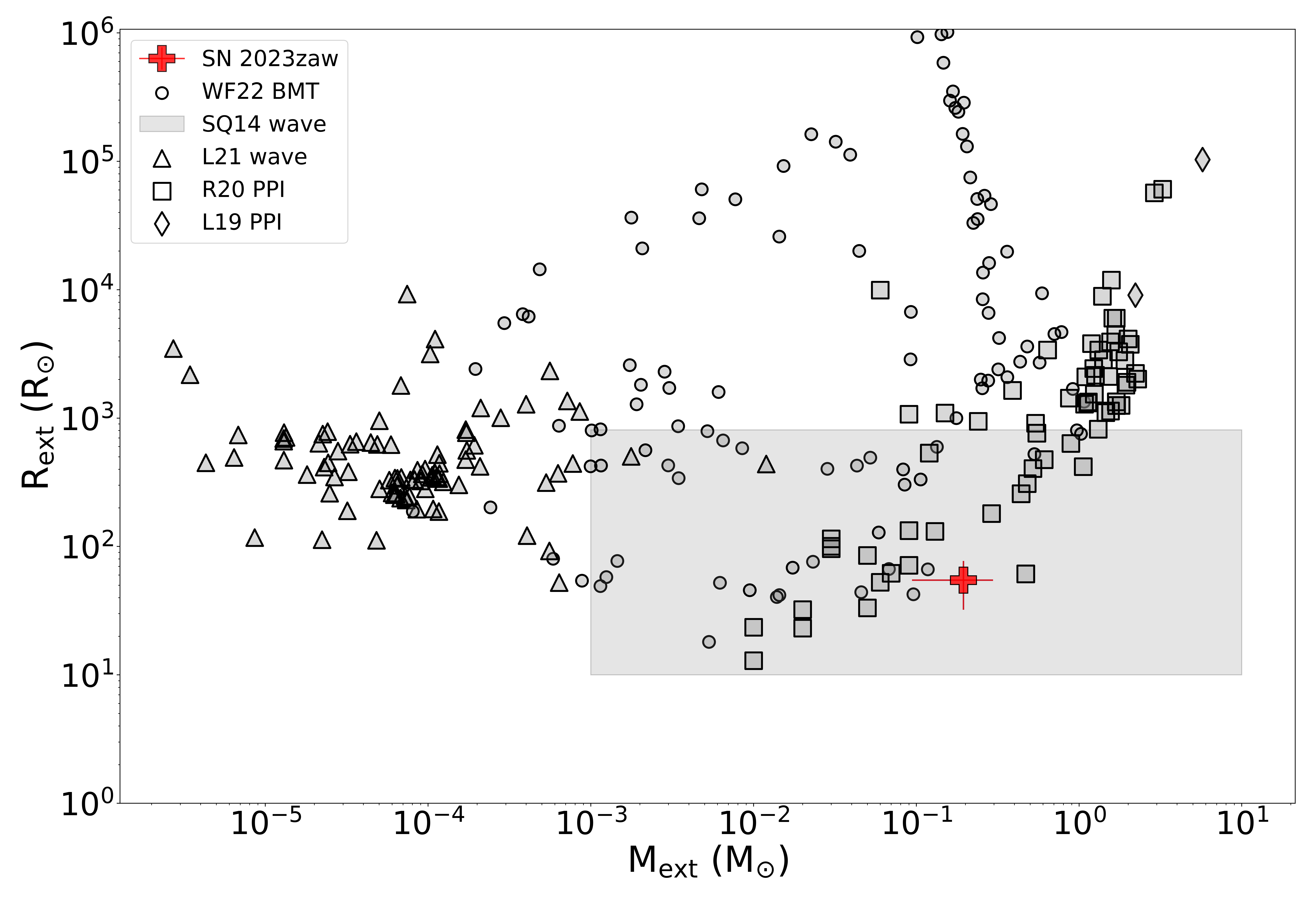}
    \caption{\textit{Left:} Comparison of the envelope properties of SN 2023zaw with theoretical models for \textit{bound} stellar material $-$ binary and single star models from \citet{Wu2022b} (WF22), \citet{Laplace2020} (L20) and \citet{Tauris2015} (T15). \textit{Right:} Comparison of the the envelope properties of SN 2023zaw with theoretical models for \textit{unbound} stellar material $-$ late-time binary mass transfer \citep[BMT;][]{Wu2022b}, wave-driven mass loss \citep{Leung2021b, Shiode2014}, pulsation-pair instability driven  mass loss \citep[PPI;][]{Renzo2020, Leung2019}.}
    \label{fig:bound}
\end{figure}

We compare the envelope properties of SN 2023zaw with USSNe candidates and Type Ibn SNe in Figure \ref{fig:envelope}. We compare the envelope properties of SN 2023zaw with theoretical models for bound and unbound stellar material in Figure \ref{fig:bound}.

\addtocounter{Afigure}{1}
\addtocounter{Atable}{1}

%\section{Comparison with other models}
%\begin{figure*}
%    \centering
%    \includegraphics[width=8.5cm]{23zaw_r_models_lessNi.png}\includegraphics[width=8.5cm]{23zaw_bol_models.png}
%    \caption{Comparison of the bolometric luminosity of SN 2023zaw with theoretical models for `.Ia' SNe \citep{Shen2010} and nickel-free SNe \citep{Kleiser2018}.}
%    \label{fig:modelcompare}
%\end{figure*}

%\section{Arnett fit without host-extinction}
%remove this later

%\begin{figure}
%    \centering
%    \includegraphics[width=8.5cm]{23zaw_arnett_fit.png}
%    \caption{Before host extinction correction. No SC; only Ni-powered}
%    \label{fig:arnettfit}
%\end{figure}

\setcounter{Atable}{0}
\setcounter{Afigure}{0}

\addtocounter{Afigure}{1}
\addtocounter{Atable}{1}

\section{Priors and Corner plots}

\begin{figure}[H]
    \centering
    \includegraphics[width=8.5cm]{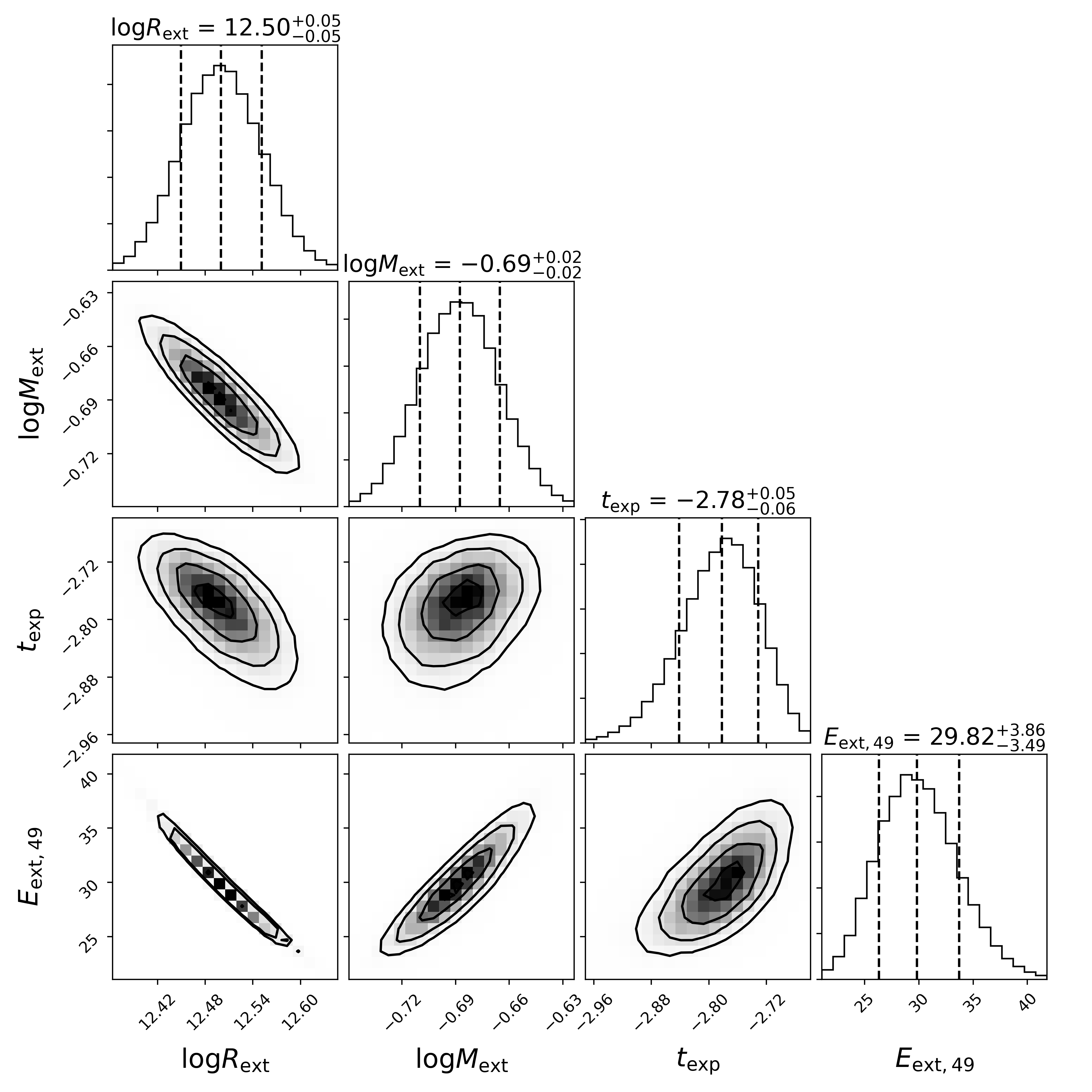}\includegraphics[width=8.5cm]{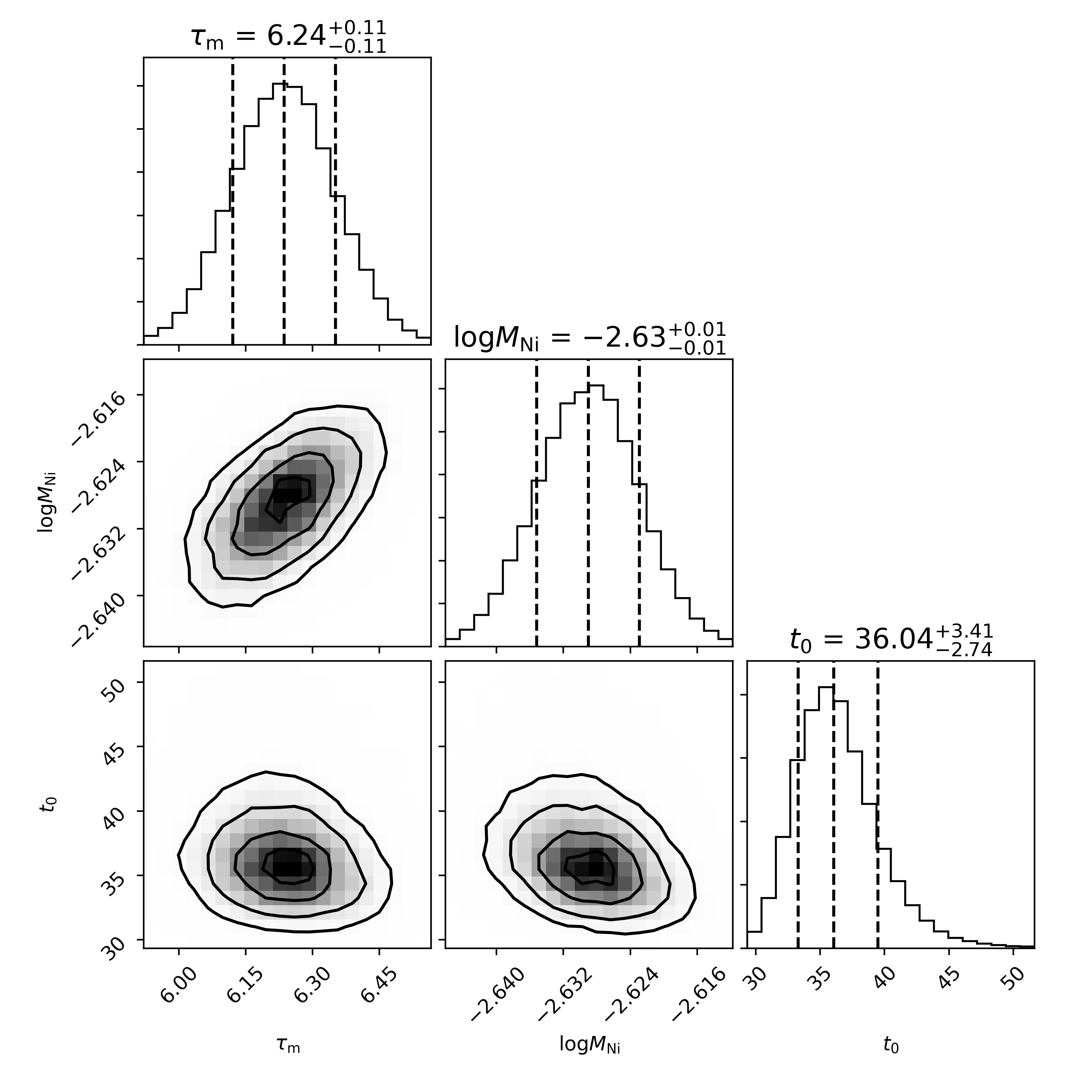}
    \caption{Corner plots showing the posterior constraints of the model parameters in the shock-cooling emission (left) and radioactive decay (right) models described in Section \ref{sec:scfit}.}
    \label{fig:scbcorner}
\end{figure}

The priors and corner plots for shock-cooling emission and radioactive decay model fitting are shown in Tables \ref{table:scbpriors}, \ref{table:nipriors} and Figure \ref{fig:scbcorner}.

\begin{table}[H] 
\begin{center} 
\caption{Priors used for shock-cooling emission model fitting described in Section \ref{sec:scfit}.} 
\begin{tabular}{cc} 
\hline 

Parameter &  Prior \\ \hline

${\rm log}R_{\rm ext}$ & $\mathcal{U}(-5, 25)$ \\ 
${\rm log}M_{\rm ext}$ & $\mathcal{U}(-4, 1)$\\
$E_{\rm ext, 49}$  & $\mathcal{U}(0.1,100)$ \\ \hline

\end{tabular}  \label{table:scbpriors} 
\end{center} 
\end{table}

\addtocounter{Atable}{1}

\begin{table}[H] 
\begin{center} 
\caption{Priors used for nickel decay model fitting described in Section \ref{sec:scfit}.} 
\begin{tabular}{cc} 
\hline 

Parameter &  Prior \\ \hline

$\tau_{\rm m}$  & $\mathcal{U}(1, 20)$ \\ 
${\rm log}M_{\rm Ni}$ & $\mathcal{U}(-4, 0)$\\
$t_0$  & $\mathcal{U}(2,  100)$ \\ \hline

\end{tabular}  \label{table:nipriors} 
\end{center} 
\end{table}

\end{document}